\documentclass[11pt,a4paper]{article}

\usepackage[T1]{fontenc}
\usepackage[utf8]{inputenc}
\usepackage[english]{babel}

\DeclareUnicodeCharacter{00A0}{~}   
\DeclareUnicodeCharacter{202F}{\,}  
\DeclareUnicodeCharacter{2217}{\ensuremath{\ast}} 
\DeclareUnicodeCharacter{2013}{-}  
\DeclareUnicodeCharacter{2014}{--} 

\usepackage[
  left=1.5cm,
  right=1.5cm,
  top=2cm,
  bottom=2cm,
  footskip=1cm
]{geometry}
\usepackage[parfill]{parskip}
\usepackage{microtype}

\usepackage{amsmath}
\usepackage{newpxtext,newpxmath}

\usepackage{graphicx}
\usepackage{booktabs}
\usepackage{longtable}
\usepackage{pdflscape}
\usepackage[table,dvipsnames]{xcolor}
\usepackage[font=small,labelfont=bf]{caption}

\graphicspath{{./fig/}}

\usepackage[affil-it]{authblk}

\usepackage[authoryear,round]{natbib}

\definecolor{myblue}{RGB}{0,90,150}
\usepackage[colorlinks=true, linkcolor=myblue,citecolor=myblue, urlcolor=myblue]{hyperref}
\title{Joint Bayesian models for validating spatial health-event databases against a gold standard: separating global and local discrepancies}

\author[1,2]{Mathias Brugel}
\author[1]{Florine Kempf}
\author[1]{Camille Ternynck}
\author[3]{Marta Blangiardo}
\author[1]{Michaël Génin}

\affil[1]{Univ Lille, CHU Lille, ULR 2694 - METRICS: Évaluation des technologies de santé et des pratiques médicales, F-59000 Lille, France}
\affil[2]{Department of Digestive Oncology, Hepatology and Gastroenterology, Centre Hospitalier de La Côte Basque, Bayonne, France}
\affil[3]{MRC Centre for Environment \& Health, Department of Epidemiology and Biostatistics, School of Public Health, Imperial College London, London, United Kingdom}

\date{}

\begin{document}

\maketitle

\section*{Abstract}

The reuse of medico-administrative and synthetic spatial data may overcome some limitations of population-based registries, provided rigorous validation is performed. However, no tool exists to spatially validate a candidate-for-reuse database (CFRD) against a gold standard (GS). We propose a Bayesian framework for two-dimensional (global and local) map-to-map validation of spatial health-event databases.
We consider an error-model family (random [REM] and structured [SEM]) in which the CFRD is modelled as a departure from the GS. Both are compared with a shared component model (SCM). Global disagreement is assessed using the database-specific intercept difference ($RR_{\mathrm{global}}$), while local disagreement is measured by the exceedance probability of the database-specific error term. Disturbance scenarios included null, uniform, clustered, and random perturbations in the CFRD. Sensitivity, specificity, false detection rate, and Matthews Correlation Coefficient assessed detection performance.
$RR_{\mathrm{global}}$ accurately recovered map-wide shifts across all models and scenarios. REM and SEM behaved were both sensitive and specific to local discrepancies. SCM was more conservative.
Applied to Crohn's disease data from the EPIMAD registry and a CFRD, all models reached the same conclusion: the CFRD reproduced global and local spatial structures with an overall signal about 7\% lower.
Extensions to other outcome distributions, spatio-temporal models and calibration constitute natural next steps.

\textit{Keywords:} data reuse; spatial database validation; Bayesian hierarchical models; disease mapping; shared component model.

\section{Introduction}


Population-based registries are epidemiological gold-standard data but remain geographically limited, costly to maintain over time and not widely accessible \citep{the_lancet_cancer_2025}. The secondary use of data enables new questions to be addressed from existing data, at lower cost and within shorter timeframes \citep{skovgaard_review_2019}. In spatial epidemiology, health data reuse is especially relevant as it can substantially extend the spatial coverage of a study and facilitate the investigation of its causes. Medico-administrative data are a potentially valuable data source, but remain poorly spatially validated \citep{remontet2008possible,BOUSSAT2024202383}. Moreover, the advent of spatial synthetic data opens new applications. Once generated, one must assess whether the synthetic version adequately reproduces the spatial signal of the original source without materially distorting its geographical structure \citep{Paiva_2014,QUICK201837,Quick_2021}. Such reuse should remain conditional on rigorous validation against an appropriate reference source.

In this setting, the comparison is paired and asymmetric. Both data sources are observed on the same areal support and are intended to describe the same underlying spatial phenomenon, area by area, making them paired. Moreover, one source is treated as the reference, or gold standard, whereas the second source is evaluated as a candidate reusable database, introducing asymmetry in the comparison. Therefore, validation of spatial databases versus a gold-standard goes beyond generic spatial pattern comparison. The objective is to determine if a database adequately reproduces both common \emph{global} latent spatial structure and \emph{local} disagreement while accounting for spatial dependence, count-data uncertainty and the asymmetric nature of the relationship between both databases. A useful statistical framework should answer to this complex issue.

Existing methodological literature only partially addresses this objective. Global and local Moran's~I and Geary's~c were initially developed to characterise autocorrelation in a single or a set of maps \citep{anselin_local_1995}. Lee's bivariate index combining Pearson's correlation and Moran's~I is another notable example \citep{lee_2001, lin_comparison_2023}. These metrics rely on \textit{t}-tests, weighted kappa coefficients, and related agreement summaries \citep{levine_2009}. They describe spatial co-structure or overall agreement, but remain essentially descriptive as they do not explicitly model areal count uncertainty, encode an asymmetric reference-versus-candidate comparison, nor decompose into shared spatial signal, global shift, and local discrepancy.

Spatial scan statistics were initially developed for cluster detection, and can be repurposed to local disagreement assessment. They search over a collection of candidate windows and identify those maximising a discrepancy or likelihood-ratio criterion \citep{kulldorf_scan_2007,kulldorff_scan_2009,cucala_multivariate_2017,cucala_multivariate_2019,Ahmed:2021aa}. They have already been used to compare two spatially structured databases, including for Crohn's disease in France \citep{Genin:2020aa}. However, they are primarily designed to identify one or several unusual clusters rather than to assess agreement over an entire territory. Moreover, their interpretation depends on the geometry of candidate windows and they are generally formulated symmetrically rather than in a reference-versus-candidate perspective. Most importantly, they do not directly distinguish map-wide shift from local residual disagreement.

Spatial point process methods have been used in this context; for instance Ripley's $K$-function, a permutation procedure comparing point patterns \citep{hahn2012}. More recent developments compare spatial intensity surfaces including kernel-based nonparametric tests, comparative evaluations of first-order procedures, and kernel mean embedding approaches for normalized intensity comparison \citep{fuentes-santos_nonparametric_2017,fuentes-santos_comparative_2023,rustamov_klosowski_2020}. These methods do not address areal count data and typically target a symmetric two-sample question of spatial similarity.

Bayesian hierarchical spatial models provide a natural basis for addressing this gap. They allow joint modelling of spatially structured outcomes, explicit representation of spatial dependence, and coherent uncertainty quantification at both global and local scales \citep{blangiardo_spatial_2015}. They have also proved useful for decomposing shared and source-specific spatial variation across related diseases or outcomes \citep{gomez-rubio_bayesian_2019}. However, to our knowledge, they have not been systematically developed and evaluated for the specific problem considered here: the validation of one spatial health-event database against another when the comparison is both \emph{paired} and \emph{asymmetric}.

The main contribution of this paper is to recast spatial database comparison as an asymmetric validation problem with distinct global and local inferential targets. We develop a principled framework for \emph{map-to-map validation} of paired spatial health-event databases against a reference source. We consider three joint Bayesian models: a random error model (REM), a structured error model (SEM), and a shared component model (SCM), as alternative latent representations of common and source-specific spatial variation. Across these models, we distinguish two complementary dimensions of disagreement: a \emph{global} discrepancy, captured by the intercept contrast between databases, and a \emph{local} discrepancy, assessed through posterior summaries derived from source-specific latent components. We evaluate the operating characteristics of this framework in a simulation study covering different perturbation structures, magnitudes, and directions, and illustrate its practical relevance by comparing Crohn's disease incidence from the French inflammatory bowel disease registry (EPIMAD), taken as the reference source, with estimates derived from the French national hospital discharge database (PMSI).

\section{Methodology}

We rely on Bayesian hierarchical spatial models to compare two spatial databases through global and local discrepancy summaries. Section~\ref{sec:models} introduces the common likelihood and notation. Section~\ref{sec:rem+sem} presents the primary asymmetric error-model family, namely the random error model (REM) and the structured error model (SEM), in which database~2 is modelled as a departure from the reference database~1. Section~\ref{sec:pp_criteria} defines the global intercept contrast, the local posterior discrepancy summaries, and the associated decision rules. Section~\ref{sec:scm} introduces the shared component model (SCM) as a symmetric benchmark comparator, and Section~\ref{sec:priors} describes the priors on hyperparameters and implementation.

\subsection{General modelling framework}
\label{sec:models}

Consider a collection of $I$ areal units indexed by $i \in \{1,\dots,I\}$ and two data sources indexed by $d \in \{1,2\}$, where database~1 is regarded as the reference source and database~2 as the source under validation. For each pair $(i,d)$, let $O_i^{(d)} \in \mathbb{N}$ denote the observed count, $E_i^{(d)} > 0$ the expected count, and $\theta_i^{(d)} > 0$ the corresponding relative risk. Conditionally on the latent relative risks, we assume the same Poisson sampling distribution for all models,
\begin{equation}
O_i^{(d)} \mid \theta_i^{(d)}
\sim
\mathcal{P}\!\left(E_i^{(d)}\theta_i^{(d)}\right),
\qquad
\log \theta_i^{(d)} = \eta_i^{(d)},
\label{eq:poisson_likelihood}
\end{equation}
where $\eta_i^{(d)}$ denotes the linear predictor.

In the paired validation setting considered here, we place expected counts on a common standardisation scale by anchoring database~2 to the reference source, so that $E_i^{(2)} = E_i^{(1)}$ for $i=1,\dots,I$. This ensures that both databases are compared against the same reference map, rather than through differences induced by source-specific re-standardisation.

All models considered below share the likelihood in \eqref{eq:poisson_likelihood} and differ only through the specification of the linear predictors and the decomposition of the latent spatial terms.

\subsection{Asymmetric error-model family}
\label{sec:rem+sem}

The error-model family treats database~1 as a gold-standard representation of the underlying spatial risk surface and models database~2 as a perturbation of that reference.

Let $W=(w_{ij})_{1 \le i,j \le I}$ be the binary neighbourhood matrix induced by Queen contiguity, with $w_{ij}=1$ if areal units $i$ and $j$ share at least one boundary point and $w_{ij}=0$ otherwise, for $i\neq j$. Let $Q(W)$ denote the intrinsic conditional autoregressive (ICAR) precision matrix associated with $W$ \citep{besag_bayesian_1991}. Spatially structured latent fields are represented through the Besag--York--Molli\'e 2 (BYM2) parametrisation proposed by \cite{riebler_pcprior_2016}. For a latent field $\mathbf{b}=(b_1,\dots,b_I)^\top$, we write $\mathbf{b} \sim \mathrm{BYM2}(W,\tau_b,\phi_b)$ if
\[
\mathbf{b}
=
\tau_b^{-1/2}
\left(
\sqrt{\phi_b}\,\mathbf{u}^{\ast}
+
\sqrt{1-\phi_b}\,\mathbf{v}^{\ast}
\right),
\]
where $\mathbf{u}^{\ast}$ is a scaled ICAR component associated with $Q(W)$, $\mathbf{v}^{\ast} \sim \mathcal{N}_I(\mathbf{0},\mathbf{I}_I)$, $\tau_b>0$ is a global precision parameter, and $\phi_b \in [0,1]$ controls the proportion of marginal variance attributable to the spatially structured component. Throughout, the ICAR component is scaled so that the BYM2 parametrisation admits an interpretable marginal variance decomposition \citep{riebler_pcprior_2016}. Standard sum-to-zero constraints are implicitly imposed on the structured components to ensure identifiability.

Let $\mathbf{s}=(s_1,\dots,s_I)^\top$ denote a latent shared spatial field and let $\mathbf{e}=(e_1,\dots,e_I)^\top$ denote a database-2-specific discrepancy field. The linear predictors are
\begin{equation}
\eta_i^{(1)} = \alpha^{(1)} + s_i,
\qquad
\eta_i^{(2)} = \alpha^{(2)} + \delta s_i + e_i,
\qquad i=1,\dots,I,
\label{eq:error_models}
\end{equation}
where $\alpha^{(1)}$ and $\alpha^{(2)}$ are database-specific intercepts and $\delta \in \mathbb{R}$ is a scaling parameter controlling the contribution of the shared spatial field to database~2. To identify a common reference surface and to interpret $e_i$ as a local residual discrepancy of database~2 relative to database~1, we fix $\delta=1$. Under this constraint, the two databases share the same latent spatial pattern $\mathbf{s}$, while departures of database~2 from the reference are encoded by the intercept contrast $\alpha^{(2)}-\alpha^{(1)}$ and the residual field $\mathbf{e}$.

The common latent spatial field is modelled as
\[
\mathbf{s} \sim \mathrm{BYM2}(W,\tau_s,\phi_s).
\]
The distinction between the two error models lies in the stochastic specification of the discrepancy field. In the random error model (REM), local departures are assumed conditionally independent across areas,
\[
e_i \overset{\mathrm{i.i.d.}}{\sim} \mathcal{N}(0,\tau_e^{-1}),
\qquad i=1,\dots,I.
\]
Thus, the REM accommodates local deviations of database~2 from the common spatial surface without imposing residual spatial dependence.

In the structured error model (SEM), the discrepancy field is instead allowed to exhibit residual spatial structure,
\[
\mathbf{e} \sim \mathrm{BYM2}(W,\tau_e,\phi_e).
\]
The SEM therefore captures residual discrepancies that remain spatially coherent after adjustment for the shared component.

As the REM isolates purely area-specific departures without borrowing information across neighbouring areas, comparing both models therefore provides a sensitivity check on whether local conclusions depend on residual spatial smoothing.

\subsection{Global and local posterior discrepancy summaries}
\label{sec:pp_criteria}

Global discrepancy is obtained by the difference of database-specific intercepts,
\begin{equation}
\Delta = \alpha^{(2)}-\alpha^{(1)},
\label{eq:global_delta}
\end{equation}
which summarises the average log-scale shift between the two databases. On the relative-risk scale, the associated multiplicative contrast is
\begin{equation}
RR_{\mathrm{global}} = \exp(\Delta).
\label{eq:RR_global}
\end{equation}
Thus, $RR_{\mathrm{global}}>1$ indicates that, on average, database~2 lies above database~1, whereas $RR_{\mathrm{global}}<1$ indicates a global downward shift. This measure reflects the overall difference between the two databases and should be interpreted as a model-based multiplicative contrast, not as an average of area-level relative-risk ratios. It is intended to capture map-wide discrepancies that may remain invisible to purely local decision rules.

For local discrepancy summaries in the asymmetric error-model family, let $\mathcal{Y}$ denote the full observed data and define
\[
\psi_i=e_i,
\qquad i=1,\dots,I.
\]
Under both the REM and the SEM, $\psi_i$ has a direct interpretation as the residual local departure of database~2 from database~1 after accounting for the shared spatial field. Inference is therefore based on the marginal posterior distribution
\[
\pi_i(\psi)=\pi(\psi_i=\psi\mid\mathcal{Y}).
\]

The first posterior functional is the \emph{null-referenced exceedance probability} (NREP),
\begin{equation}
\mathrm{NREP}_i
=
\Pr(\psi_i>0 \mid \mathcal{Y})
=
\int_0^{+\infty} \pi_i(\psi)\, d\psi.
\label{eq:NREP_def}
\end{equation}
Because $\psi_i$ is defined on the log-relative-risk scale, the event $\psi_i>0$ is equivalent to $\exp(\psi_i)>1$. The quantity $\mathrm{NREP}_i$ therefore quantifies the posterior evidence that, after adjustment for the shared spatial component, database~2 exhibits a positive local residual departure relative to database~1. Although \eqref{eq:NREP_def} is one-sided, the associated decision rule is two-sided. For fixed constants $0<\ell<u<1$, area $i$ is declared discrepant whenever $\mathrm{NREP}_i \le \ell$ or $\mathrm{NREP}_i > u$.

A second posterior functional is introduced to distinguish genuinely local departures from a possible territory-wide shift. This is motivated by the structure of BYM2 latent fields. Because the structured components are subject to sum-to-zero constraints, broad positive departures over part of the territory must be compensated elsewhere by negative values, even when the scientific signal is primarily global rather than local. As a result, direct thresholding relative to zero may exaggerate contrast induced by the centring constraints of the latent field. To mitigate this effect, we define a \emph{robustly centred exceedance probability} (RCEP) relative to a data-driven global centre.

Let $m_i = \mathrm{Med}(\psi_i \mid \mathcal{Y})$ be the posterior median of $\psi_i$ in area $i$, and define the global centre by
\[
c^\star = \mathrm{median}\{m_1,\dots,m_I\}.
\]
The quantity $c^\star$ is a robust estimator of the central tendency of the latent discrepancy field. Its use is intended to separate a broad, approximately homogeneous shift in the discrepancy surface from genuinely atypical local behaviour. The use of such a robust centre was not intended to define a uniquely canonical Bayesian discrepancy functional, but rather to provide an operational and interpretable way to separate broad background shift from locally atypical behaviour in centred latent spatial models.

Let $\varepsilon>0$ denote a tolerance radius on the log-relative-risk scale. The \emph{robustly centred exceedance probability} is defined by
\begin{equation}
\mathrm{RCEP}_i
=
\Pr\!\big(|\psi_i-c^\star|>\varepsilon \mid \mathcal{Y}\big)
=
1-
\int_{c^\star-\varepsilon}^{c^\star+\varepsilon}
\pi_i(\psi)\, d\psi.
\label{eq:RCEP_def}
\end{equation}
Equivalently, on the relative-risk scale, $\mathrm{RCEP}_i = \Pr\!\left( \exp(\psi_i) \notin \left[\exp(c^\star-\varepsilon),\,\exp(c^\star+\varepsilon)\right] \middle| \mathcal{Y} \right). $ Thus, $\mathrm{RCEP}_i$ measures the posterior evidence that the local discrepancy in area $i$ lies outside a robust equivalence band centred on the overall discrepancy level rather than on the null value. For a threshold $\tau \in (0,1)$, the centred decision rule declares area $i$ discrepant whenever $\mathrm{RCEP}_i > \tau.$

The distinction between \eqref{eq:NREP_def} and \eqref{eq:RCEP_def} is fundamental. The NREP assesses whether the local discrepancy is credibly away from the null reference value, whereas the RCEP assesses whether it is atypical relative to the global centre of the discrepancy field. The latter is therefore especially relevant when one wishes to distinguish localised disagreement from broad contrast patterns partly induced, or amplified, by the centring structure of the latent model. Their comparative operating characteristics are assessed in the simulation study described in Section~\ref{sec:simulation}.

The quantity $c^\star$ is estimated from the fitted model and subsequently treated as fixed in \eqref{eq:RCEP_def}. Accordingly, the RCEP is a plug-in posterior functional rather than a fully Bayesian quantity integrating over the posterior uncertainty of $c^\star$.

\label{sec:posterior_rules}

For each fitted model, inference on the global contrast was summarised by the posterior mean of $\Delta$ together with its $95\%$ credible interval. The same summaries were reported on the relative-risk scale through the posterior mean of $RR_{\mathrm{global}}=\exp(\Delta)$ and the corresponding $95\%$ credible interval obtained from the posterior distribution of $\Delta$ by exponentiation.

For local classification under the asymmetric error-model family, the two posterior decision rules introduced in Section~\ref{sec:pp_criteria} were applied to the marginal posterior distribution of $\psi_i=e_i$.

The first rule was based on the null-referenced exceedance probability, $\mathrm{NREP}_i=\Pr(\psi_i>0\mid\mathcal{Y})$, used through its two-sided formulation. For threshold pairs $(\ell,u)\in\{(0.2,0.8),\,(0.05,0.95),\,(0.025,0.975)\}$, area $i$ was declared discrepant whenever $\mathrm{NREP}_i \le \ell$ or $\mathrm{NREP}_i > u$. This rule identifies areas for which the posterior mass is sufficiently concentrated away from the null reference value, regardless of direction.

The second rule was based on the robustly centred exceedance probability, $\mathrm{RCEP}_i=\Pr\!\bigl(|\psi_i-c^\star|>\varepsilon \mid \mathcal{Y}\bigr)$. In practice, $c^\star$ was computed as the median of the area-specific posterior medians of $\psi_i$, and the tolerance parameter was fixed at $\varepsilon=\log(1.10)$, corresponding to a $10\%$ equivalence band on the relative-risk scale. For thresholds $\tau\in\{0.8,\,0.9,\,0.95\}$, area $i$ was declared discrepant whenever $\mathrm{RCEP}_i>\tau$.

The application of these posterior decision rules to the SCM benchmark requires a model-specific definition of the operational residual summary and is described in the next section.

\subsection{Shared component model as benchmark}
\label{sec:scm}

As a benchmark, we also consider a shared component model (SCM), a standard formulation for jointly modelling related spatial outcomes. Unlike the REM and SEM, this model is symmetric at the latent level and does not explicitly represent database~2 as a departure from database~1.

The linear predictors are
\begin{equation}
\eta_i^{(1)} = \alpha^{(1)} + \delta^{(1)} h_i + D_i^{(1)},
\qquad
\eta_i^{(2)} = \alpha^{(2)} + \delta^{(2)} h_i + D_i^{(2)},
\qquad i=1,\dots,I,
\label{eq:SCM}
\end{equation}
where $\alpha^{(1)}$ and $\alpha^{(2)}$ are source-specific intercepts, $\mathbf{h}=(h_1,\dots,h_I)^\top$ is a latent field shared by the two databases, $\delta^{(1)}$ and $\delta^{(2)}$ are source-specific scaling parameters, and $\mathbf{D}^{(1)}=(D_1^{(1)},\dots,D_I^{(1)})^\top$ and $\mathbf{D}^{(2)}=(D_1^{(2)},\dots,D_I^{(2)})^\top$ are database-specific residual components.

The latent fields are assigned the priors
\[
\mathbf{h} \sim \mathrm{BYM2}(W,\tau_h,\phi_h),
\qquad
\mathbf{D}^{(1)} \sim \mathrm{BYM2}(W,\tau_{D^{(1)}},\phi_{D^{(1)}}),
\qquad
\mathbf{D}^{(2)} \sim \mathrm{BYM2}(W,\tau_{D^{(2)}},\phi_{D^{(2)}}).
\]

For identifiability and interpretability, we fix $\delta^{(1)}=\delta^{(2)}=1$, so that
\[
\eta_i^{(1)} = \alpha^{(1)} + h_i + D_i^{(1)},
\qquad
\eta_i^{(2)} = \alpha^{(2)} + h_i + D_i^{(2)}.
\]
This specification does not constrain the sign of $h_i$ itself: the shared field may take either positive or negative values according to the posterior distribution induced by the BYM2 prior and the likelihood. Rather, fixing identical positive scaling parameters ensures that the shared component enters both linear predictors with the same orientation and on the same scale. In that sense, $\mathbf{h}$ captures the part of the spatial variation that is concordant across the two databases, while residual variation is absorbed by the database-specific components $\mathbf{D}^{(1)}$ and $\mathbf{D}^{(2)}$.

The contrast with the error-model family is substantial. In the REM and SEM, the decomposition is intrinsically asymmetric, since database~1 is privileged as the reference surface and database~2 is represented conditionally upon it through an explicit discrepancy field $\mathbf{e}$. By contrast, the SCM is symmetric at the latent level: both databases are expressed as parallel deviations around a common spatial structure. Consequently, $D_i^{(2)}$ does not, strictly speaking, represent a direct local contrast between database~2 and database~1. It represents the component of database~2 that remains unexplained after accounting for the shared field $\mathbf{h}$.

For benchmark purposes, we therefore defined the SCM-specific operational residual summary as
\[
\psi_i^{\mathrm{SCM}} = D_i^{(2)}.
\]
The NREP and RCEP functionals were then computed using $\psi_i^{\mathrm{SCM}}$ with the same thresholds as those used for the REM and SEM. These SCM-based summaries should not be interpreted as posterior probabilities attached to a direct source-to-source discrepancy. Rather, they provide a symmetric-model benchmark for detecting residual spatial structure in the candidate database after accounting for the shared component.

We did not use the explicit residual contrast $D_i^{(2)}-D_i^{(1)}$ as the primary SCM discrepancy summary, because this contrast combines two source-specific residual components estimated symmetrically and therefore does not preserve the reference-versus-candidate interpretation targeted by the validation problem.

\subsection{Priors on hyperparameters and implementation}
\label{sec:priors}

Weakly informative regularising priors were assigned to the hyperparameters governing the latent fields. For all BYM2 precision parameters, namely those associated with $\mathbf{s}$, with $\mathbf{e}$ under the SEM, and with $\mathbf{h}$, $\mathbf{D}^{(1)}$, and $\mathbf{D}^{(2)}$ under the SCM, we used penalised complexity priors in the sense of \cite{riebler_pcprior_2016}, parameterised through the tail event $\Pr\!\left(\tau^{-1/2}>1\right)=0.01$. For the BYM2 mixing parameters, we used PC priors favouring neither the structured nor the unstructured component a priori, corresponding to $\Pr(\phi<0.5)=0.5$.
For the REM, where the discrepancy field is purely unstructured, $e_i \overset{\mathrm{i.i.d.}}{\sim}\mathcal{N}(0,\tau_e^{-1})$, we assigned a weakly informative log-Gamma prior with parameters $(1,\,0.00005)$ to $\tau_e$.
Overall, these priors provide regularisation while allowing the data to drive the posterior decomposition between shared spatial signal and source-specific departures.

All models were fitted in \texttt{R} using the Integrated Laplace Nested Approximation (INLA) framework \citep{rue2009approximate}. INLA provides numerical approximations to the marginal posterior distribution of each latent effect. Posterior decision functionals were computed from the marginal posterior distributions returned by INLA.

\section{Simulation study}
\label{sec:simulation}

\subsection{Simulation design}
\label{sec:data-generation}

The simulation study was conducted on the 100 areal units of the \texttt{sf::nc} dataset, corresponding to counties in North Carolina (USA). Let $\mathcal{A}=\{1,\dots,I\}$ denote the study region, with $I=100$. Spatial dependence was encoded through a binary Queen-contiguity matrix $W$, and the associated graph Laplacian was denoted by $R$.

Data were generated using a Leroux model \citep{leroux_estimation_2000}. By using a Leroux model instead of a BYM2, overly favourable self-generating model assessment was avoided.
Let $\mathbf{Z}=(Z_1,\dots,Z_I)^\top$ denote a latent Gaussian spatial field generated from a Leroux model \citep{leroux_estimation_2000},
\[
\mathbf{Z} \sim \mathcal{N}_I\!\left(
\mathbf{0},
\;
\sigma^2\bigl(\rho R + (1-\rho)I_I\bigr)^{-1}
\right),
\]
where $\sigma^2>0$ is a global scale parameter controlling the variability of the latent spatial field and $\rho\in[0,1]$ controls spatial dependence. For each area $i\in\mathcal{A}$, the reference database (database~1) was generated according to
\[
\lambda_i^{(1)}=\exp\!\bigl(\log(p)+\log(pop_i)+Z_i\bigr),
\qquad
O_i^{(1)} \mid Z_i \sim \mathcal{P}\!\bigl(\lambda_i^{(1)}\bigr),
\]
where $pop_i$ denotes the population at risk. Expected counts for the subsequent analysis model were then obtained by indirect internal standardisation,
\begin{equation}
E_i^{(1)}
=
pop_i
\frac{\sum_{j=1}^{I} O_j^{(1)}}{\sum_{j=1}^{I} pop_j}.
\label{eq:E_reg}
\end{equation}
The comparison database (database~2) was generated from the same latent field after introducing area-specific perturbation factors $r_i$:
\[
\lambda_i^{(2)}=\exp\!\bigl(\log(p)+\log(pop_i)+Z_i+\log(r_i)\bigr),
\qquad
O_i^{(2)} \mid Z_i \sim \mathcal{P}\!\bigl(\lambda_i^{(2)}\bigr).
\]
As in the motivating application, expected counts in database~2 were set equal to those of the reference source, that is, $E_i^{(2)} = E_i^{(1)}$ for $i=1,\dots,I$.

The main simulation design focused on the spatial structure of the perturbation, its magnitude, its direction, and, for random scenarios, the proportion of modified areas. In addition, sensitivity analyses were conducted by varying the baseline incidence level ($p = 0.01, 0.001,$ and  $0.0001$) and the spatial correlation parameter of the latent Leroux field ($\rho = 0.25, 0.50,$ and  $0.75$), while keeping the other data-generating parameters unchanged. These incidence levels corresponded approximately to mean numbers of incident cases per spatial unit of 364, 36, and 3.6, respectively, in both databases. Further sensitivity analyses were conducted for the tolerance parameter $\epsilon$ used in the robustly centred exceedance probability, considering $\epsilon=\log(1.05)$, $\log(1.10)$, and $\log(1.20)$. Because $\epsilon$ enters only the definition of the local RCEP rule and not the global intercept contrast $RR_{\mathrm{global}}=\exp(\Delta)$, this additional sensitivity analysis concerned local discrepancy detection only. These supplementary analyses were used to assess the robustness of the proposed framework to event rarity, to the strength of latent spatial dependence, and to the width of the equivalence band used for centred local decision-making; their detailed results are reported in Supplementary Figures ~\ref{fig:sup_incidence_global}--\ref{fig:sup_eps_mcc}.

For each simulation configuration, datasets were generated repeatedly until 100 successful fitted replicates were obtained for each model and decision rule. Quantities that were undefined by construction in specific scenarios were treated as missing and omitted from the corresponding summaries.

\subsection{Disturbance scenarios}
\label{sec:scenarios}

The perturbation factors $r_i$ were defined according to four scenarios. Throughout, one-sided perturbations were generated on the grid $r \in \{0.25,\,0.50,\,0.75,\,1.25,\,1.5,\,1.75,\,2.0,\,2.25,\,2.5\}$, where values below 1 correspond to downward perturbations and values above 1 to upward perturbations. Opposite perturbations scenarios were generated using the pairs $(1.5,\,0.75)$ and $(1.75,\,0.50)$.

In scenario~S1 (\emph{null}), no area was modified and $r_i = 1$ for $i=1,\dots,I$. In scenario~S2 (\emph{uniform disturbance}), all areas were modified by the same multiplicative factor, so that $r_i = r$ for $i=1,\dots,I$. In scenario~S3 (\emph{clustered disturbance}), perturbations were restricted to a fixed connected spatial subset. In one-sided clustered settings, all areas within the cluster were assigned the same factor $r$, while the remaining areas were left unchanged. Mixed clustered settings were also considered, in which one subset was perturbed upward and another disjoint subset downward. In scenario~S4 (\emph{random disturbance}), a proportion $\pi \in \{0.25,\,0.50\}$ of areas was drawn uniformly without replacement and labelled as modified. In one-sided random settings, all selected areas received the same perturbation factor $r$. In mixed random settings, each selected area was independently assigned an upward or downward perturbation with probability $0.5$. All scenarios are summarised in Table \ref{tab:simulation_scenarios}

For each simulated dataset, the true disturbance status was recorded through the indicator $Y_i$, defined by $Y_i=1$ if area $i$ was modified and $Y_i=0$ otherwise. In mixed settings, both upwardly and downwardly perturbed areas were therefore treated as truly discrepant.

\subsection{Performance measures}
\label{sec:sens+spe}

For each model, posterior decision rule, and simulation dataset, the predicted labels $\widehat{Y}_i$ were compared with the true labels $Y_i$. Performance was summarised using sensitivity, specificity, false discovery rate, and Matthews correlation coefficient (defined in Supplementary Materials \ref{supp_mat}). Taken together, these criteria characterise detection power, false-positive control, error burden among flagged areas, and overall classification performance under class imbalance. These local performance measures are most directly interpretable in scenarios preserving genuine local contrast between modified and unmodified areas. By contrast, under purely uniform perturbation, the main inferential target is global rather than local, so that near-absence of local detection should be interpreted as expected behaviour of the local rules rather than as a failure of estimation. In that sense, Scenario~S2 should primarily be read as a negative control for local discrepancy detection and as a validation setting for the global intercept contrast.

%

\subsection{Results}
\label{sec:results}

\subsubsection{Global detection across all scenarios}

\begin{figure}[htbp]
\begin{center}
\includegraphics[width=\textwidth]{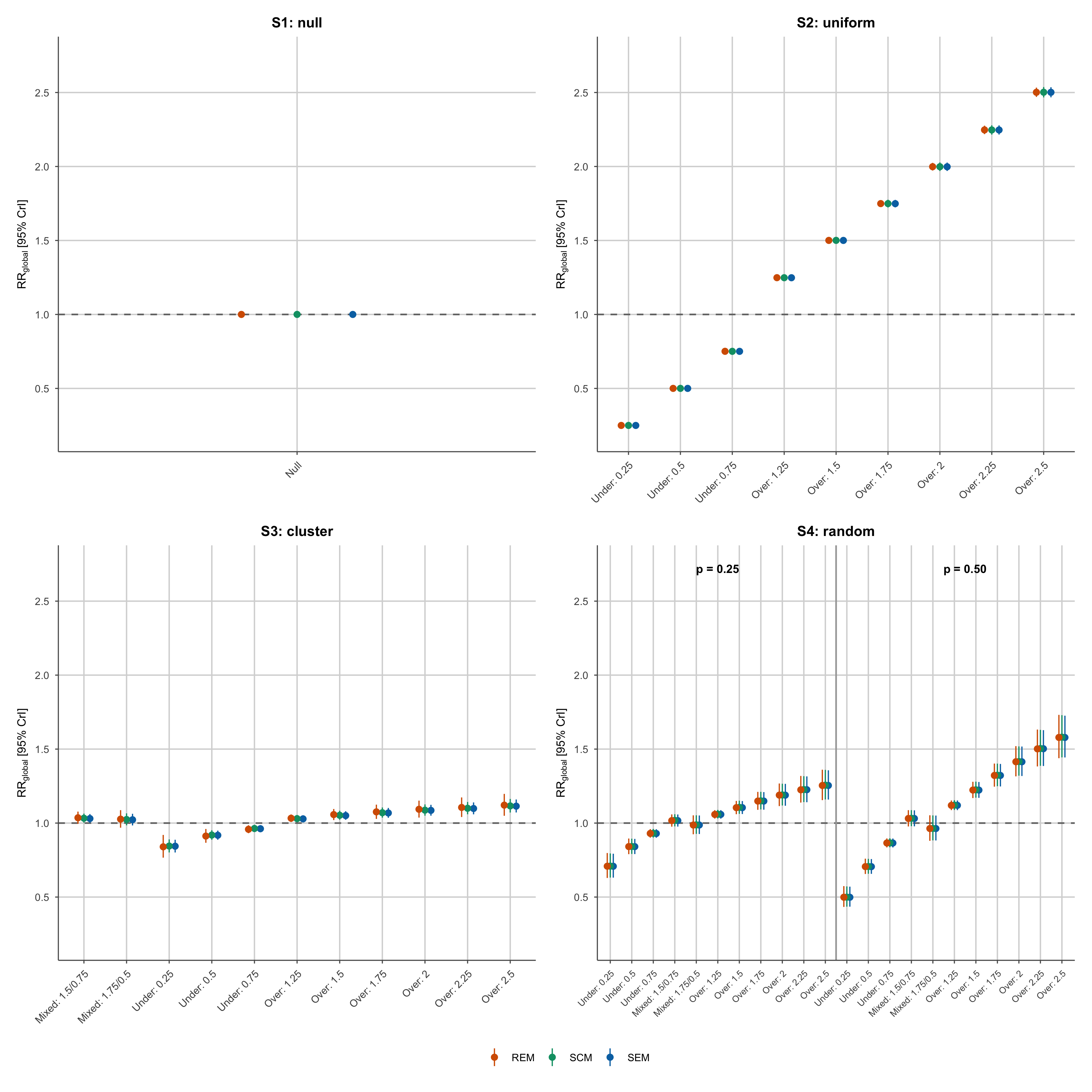}
\caption{Posterior median of the global relative-risk contrast, $RR_{\mathrm{global}}=\exp(\Delta)$, across simulation scenarios and perturbation settings, for the random error model (REM), shared component model (SCM), and structured error model (SEM), together with corresponding 95\% credible intervals (CrIs). The dashed horizontal line indicates $RR_{\mathrm{global}}=1$, corresponding to the absence of global discrepancy between the two databases. This figure highlights the ability of the intercept contrast to recover map-wide shifts while remaining close to 1 under predominantly local perturbations.}
\label{fig:global_rr}
\end{center}
\end{figure}

Figure~\ref{fig:global_rr} shows that the global intercept contrast behaved as expected across all simulation scenarios and all three models. Under the null and uniform scenario (S1 and S2), $RR_{\mathrm{global}}$ was centred almost exactly at 1 in all three models. In the clustered (S3) and random (S4) scenarios, the global contrast moved modestly away from 1 only for extreme perturbations. For S3, the maximum $RR_{\mathrm{global}}$ ranged between 1.11 to 1.12 for $r=2.5$. For S4, a similar trend was observed (upward : 1.06 for $r=1.25$ to 1.25-1.26 for $r=2.5$; downward : 0.93 for $r=0.75$ to 0.71 for $r=0.25$). For S4, the $RR_{\mathrm{global}}$ was increasing with the proportion of modified areas ($RR_{\mathrm{global}}$=1.25 for $r=2.5$ for 25\%; $RR_{\mathrm{global}}$=1.58 for $r=2.5$ for 50\%).
In mixed clustered settings, $RR_{\mathrm{global}}$ ranged between 1.02 and 1.04, reflecting the compensation between upward and downward local departures. 

\subsubsection{Local detection under the null (S1) and global perturbation (S2) scenario}
Under S1, all methods were well calibrated. Specificity was equal or extremely close to 1 for all models and thresholds, indicating an almost complete absence of false local detections when the two maps were identical. Detailed scenario-specific results for S1 are shown in Supplementary Figure~\ref{fig:sup_Sc1}. In scenario~S2, none of the models or local decision rule (RCEP or NREP) wrongly detected discrepancies. This behaviour is entirely consistent and shows the duality of the detection ability. The corresponding scenario-specific results are displayed in Supplementary Figure~\ref{fig:sup_Sc2}.

\subsubsection{Local detection under clustered perturbations (S3)}

\begin{figure}[htbp]
\begin{center}
\includegraphics[width=\textwidth]{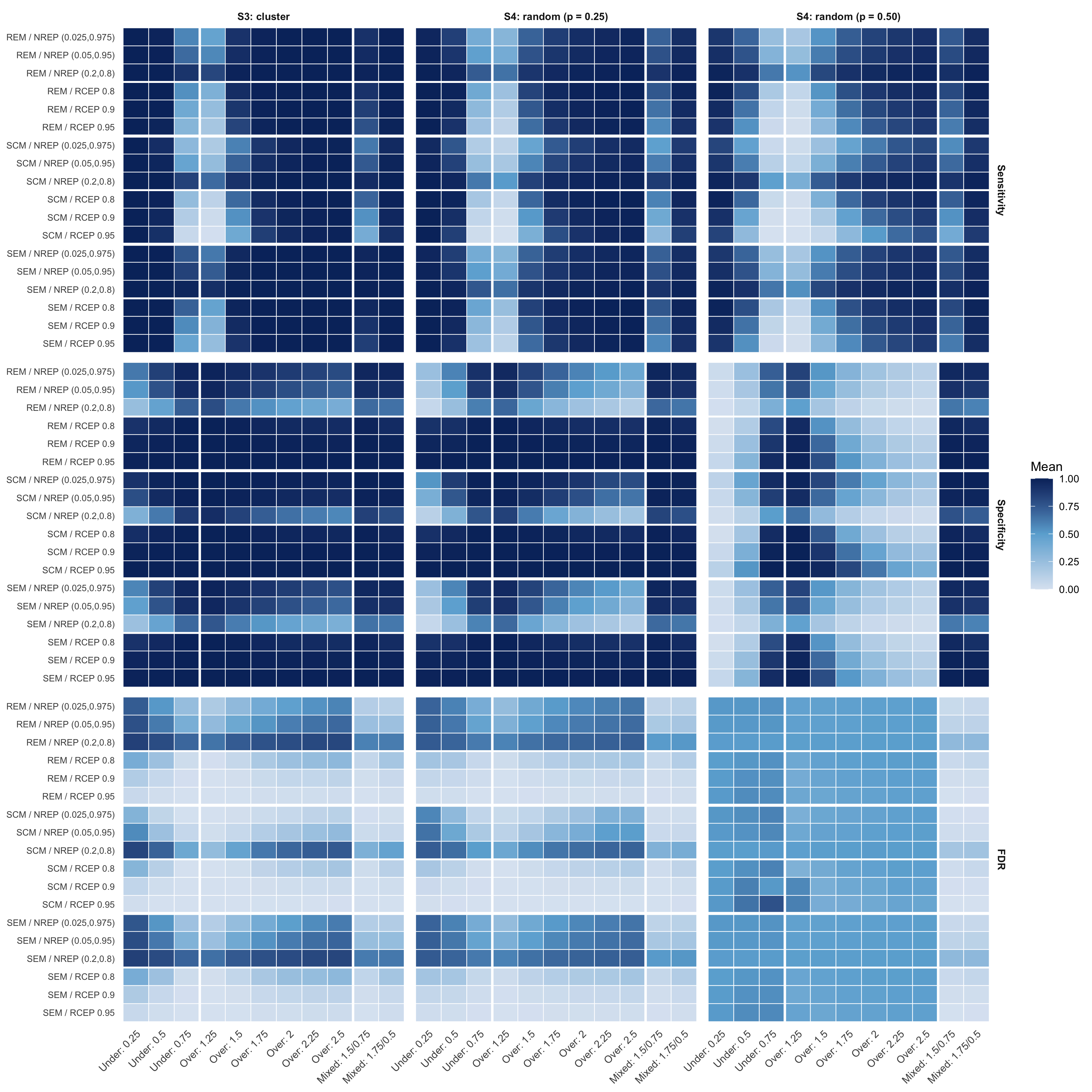}
\caption{Mean local classification performance across simulation settings for clustered perturbations (S3) and random perturbations (S4), summarised by sensitivity, specificity, and false discovery rate. Results are shown for the REM, SCM, and SEM under the null-referenced exceedance probability (NREP) and the robustly centred exceedance probability (RCEP), with several decision thresholds. Darker shading indicates higher values. Together, these heatmaps summarise how local detection depends on perturbation structure, model choice, and posterior decision rule.}
\label{fig:perf_heatmap}
\end{center}
\end{figure}

\begin{figure}[htbp]
\begin{center}
\includegraphics[width=\textwidth]{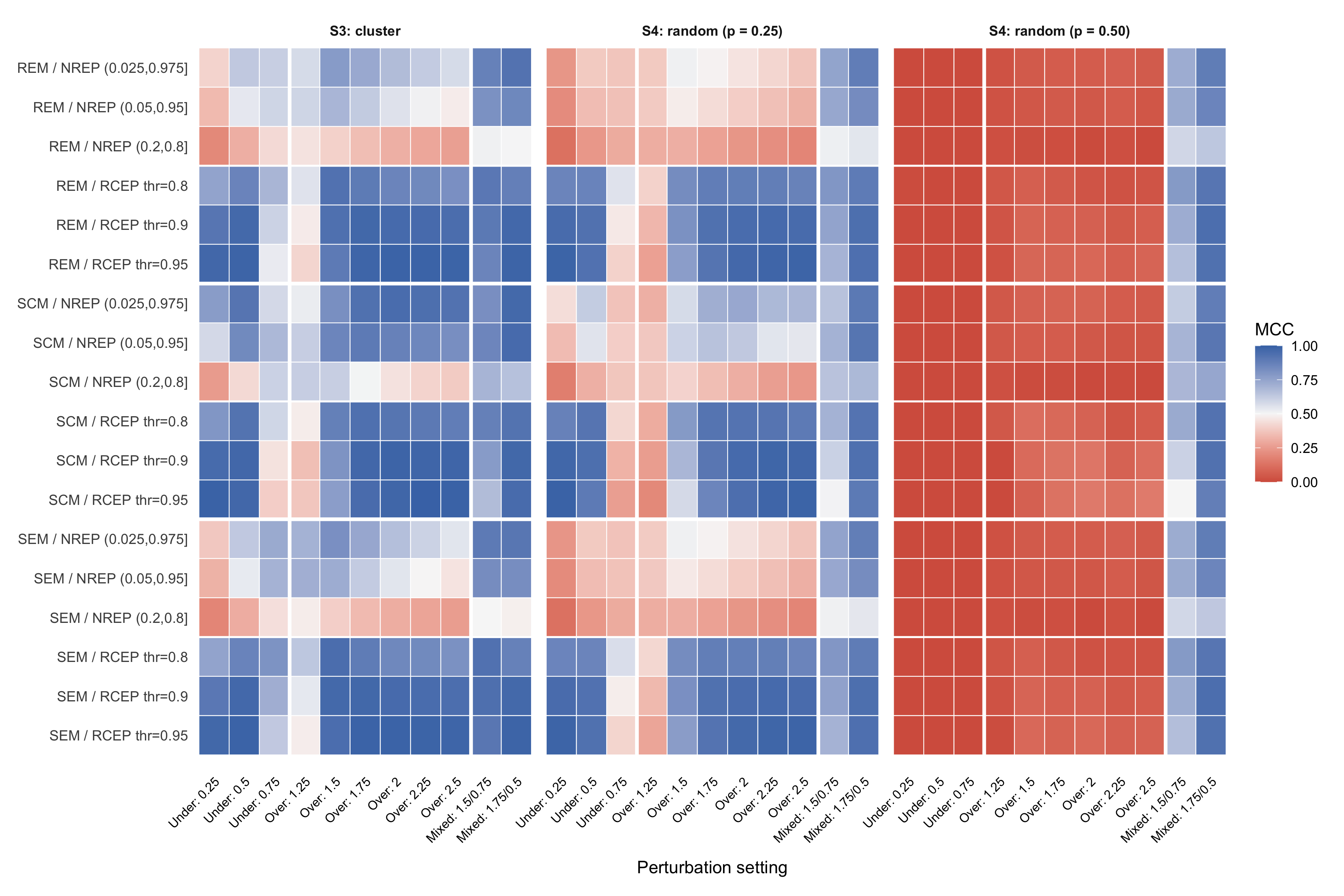}
\caption{Mean Matthews correlation coefficient (MCC) across simulation settings for clustered perturbations (S3) and random perturbations (S4), for the REM, SCM, and SEM under the null-referenced exceedance probability (NREP) and the robustly centred exceedance probability (RCEP). Higher values indicate better overall classification performance by jointly accounting for true and false positives and negatives. This figure provides a synthetic view of the trade-off between sensitivity and false-positive control across models, perturbation structures, and decision thresholds.}
\label{fig:mcc_heatmap}
\end{center}
\end{figure}

Scenario~S3 provides the most relevant and discriminating situations in which models and decision rules could be compared. SEM provided the best performance compared to REM and extensively to SCM, in terms of sensitivity, specificity and false detection rate. RCEP consistently yielded the best overall trade-off between sensitivity and false-positive control. (Figures~\ref{fig:perf_heatmap} and~\ref{fig:mcc_heatmap}). Conversely, NREP yielded lower specificity and substantially higher false discovery rates for comparable sensitivity. The only scenario in which NREP challenged RCEP sensitivity was for weak one-sided clustered perturbation, at the cost of higher false positives and MCC (for $r=1.25$ and a conservative RCEP threshold, for sensitivity: SEM:0.256 versus REM:0.332). These results support the use of the SEM with RCEP as the default local decision rule, with threshold 0.9. These settings provide the most balanced configuration in terms of performance. Detailed scenario-specific results are shown in Supplementary Figure~\ref{fig:sup_Sc3}.

\subsubsection{Local detection under random perturbations (S4, p=0.25 and p=0.5)}

Scenario~S4 provided more heterogeneous results, as performance depended strongly on both the proportion of modified areas and the preservation of local contrast (Supplementary Figures~\ref{fig:sup_Sc4_025} and~\ref{fig:sup_Sc4_05}).

When 25\% of areas were modified, results were qualitatively close to those observed in S3. SEM or REM associated with RCEP remained the most reliable configuration for local discrepancy detection, especially for moderate-to-strong perturbations (REM, for $r=2$ and RCEP threshold 0.9; sensitivity = 0.974, specificity = 0.983, FDR = 0.048, and MCC = 0.950 versus SEM, 0.974, 0.984, 0.047, and 0.951, respectively). Similar results were obtained for downward perturbations and mixed random perturbations.

Conversely, when 50\% of areas were modified in the same direction, local classification deteriorated substantially for all methods: specificity collapsed, false discovery rates approached 0.5, and MCC values became close to 0 or even negative, especially for downward perturbations (under REM, $r=0.25$, NREP sensitivity = 0.893 or 0.983 depending on the threshold pair, specificity between 0.008 and 0.037, MCC value = $-0.137$). In this setting, RCEP did not rescue performance: specificity remained extremely poor, leading again to near-zero or negative MCC values. The same pattern was observed for SEM and SCM, irrespective of the local rule.

These results indicate that once half of the territory is modified in the same direction, a sparse local anomaly transforms into a diffuse territory-wide shift, as also proven by the increase in the global signal.

Mixed random perturbations remained much easier to detect than one-sided random perturbations, even when 50\% of areas were modified. Indeed, the coexistence of upward and downward perturbations preserved a genuine local contrast structure (REM with $p=0.50$ and $(r_{+},r_{-})=(1.75,0.5)$, RCEP at threshold 0.9: sensitivity 0.960, specificity 0.980, FDR 0.021, and MCC 0.941 versus SEM at 0.961, 0.980, 0.020, and 0.942, respectively). 

Thus, the deterioration observed in dense random scenarios depends less on the number of modified areas itself than on whether the perturbation still preserves identifiable local contrast.

\subsubsection{Comparative behaviour of NREP and RCEP}

Across scenarios, NREP tended to be more liberal, more threshold-sensitive and highly sensitive, especially under clustered or sparse random alternatives compared to RCEP. This usually came with lower specificity and higher false discovery rates. By contrast, RCEP produced more stable behaviour across perturbation settings by reducing false-positive detections in scenarios where part of the apparent contrast could be induced or amplified by the centring constraints of the latent field. Outside the borderline configuration of weak one-sided perturbation, RCEP provided the most favourable and most interpretable operating characteristics. 
Overall, threshold 0.9 appeared to offer, in the settings explored here, the best compromise between sensitivity and false-positive control, whereas threshold 0.95 was slightly more conservative and threshold 0.8 slightly more liberal. Additional threshold-specific details by scenario are available in Supplementary Figures~\ref{fig:sup_Sc3}--\ref{fig:sup_Sc4_05}.

\subsubsection{Comparison between REM, SCM, and SEM}

Differences between models were generally smaller than differences between scenarios and decision rules. The two asymmetric error models behaved very similarly throughout and showed better overall performance than SCM. The benchmark SCM produced broadly concordant global conclusions but was more conservative for local detection in weak one-sided settings, consistent with its symmetric latent formulation. This shows that performance is not driven by whether the discrepancy field is represented as independent or spatially structured residual variation.

%

\subsubsection{Sensitivity analyses for incidence level, spatial correlation, and equivalence-band width}

Supplementary sensitivity analyses showed that global discrepancy estimation and local discrepancy detection were not equally affected by changes in the data-generating process or in the tuning of the centred local decision rule (Supplementary Figures~\ref{fig:sup_incidence_global}--\ref{fig:sup_eps_mcc}). 

When the baseline incidence level decreased from $p = 0.01$ (average of 364 incident cases per spatial unit) to $p = 0.0001$ (3.6 average incidence cases per spatial unit), the global discrepancy summary based on the intercept contrast remained broadly well calibrated across scenarios and models. However, uncertainty increased markedly at the rarest incidence level, especially in heterogeneous settings. 
Conversely, local discrepancy detection deteriorated rapidly as incidence decreased: sensitivity and MCC declined substantially, particularly for weak or spatially heterogeneous perturbations, while unstable or near-degenerate performance appeared in the rarest-event settings.

The spatial correlation level of the latent field had little effect on the global discrepancy summary and only modest influence on local detection performance. Across $\rho = 0.25$, $0.50$, and $0.75$, intercept contrasts were almost unchanged for all three models and all simulation scenarios. Local classification metrics varied  without any substantial change in the overall ranking of models or posterior decision rules.

Variation in the tolerance parameter $\epsilon$ of the RCEP produced the expected trade-off: smaller values yielded more liberal behaviour (higher sensitivity but lower specificity and less favourable false discovery rates), whereas larger values were more conservative. Across the settings explored here, $\epsilon=\log(1.10)$ provided the most balanced overall operating characteristics, supporting the practical relevance of the default 10\% equivalence band.

\section{Application to Crohn's disease data from a population-based registry and a hospital discharge database}
\label{sec:application}

Crohn's disease (CD) is a type of inflammatory bowel disease (IBD), displaying a marked geographical heterogeneity in both incidence and prevalence \citep{genin_space-time_2013,ng2013geographical,Genin:2020aa}. Such heterogeneity provides a natural setting in which to assess whether an external data source reproduces the spatial signal observed in a reference registry.

In France, the EPIMAD registry provides long-standing population-based surveillance of IBD incidence in northern France since the late 1980s. However, registry coverage remains geographically limited by design (9.3\% of the French territory). Extending investigations beyond the registry area therefore requires \emph{data reuse}, relying on alternative sources such as medico-administrative databases, provided that their spatial signal can be evaluated against an appropriate reference source.

In this application, CD data are obtained from two sources: (i) the EPIMAD registry, treated here as the reference source, and (ii) the French national hospital discharge database (PMSI), treated as a candidate reusable source. Data are analysed at the PMSI geographical code scale and restricted to the EPIMAD catchment area (Nord, Pas-de-Calais, Somme, and Seine-Maritime), corresponding to 574 spatial units.

The two sources do not target strictly identical epidemiological quantities. Medicoadministrative data count the number of unique hospital stays with at least one hospital stay coded for CD between 2007 and 2014 and should not be interpreted as a direct measure of incidence \citet{Genin:2020aa}. It reflects the spatial distribution of hospital-treated CD over the study period. The present application should therefore be interpreted as a comparison of spatial signal between a registry-based reference source and a hospital-based medico-administrative source, rather than as a strict comparison of identical epidemiological estimands.

Moreover, the EPIMAD registry provides CD cumulative incidence data from 1988--2014 (11,582 cases), whereas PMSI-based estimates cover 2007--2014 (10,867 cases). The longer observation window retained for EPIMAD reflects the fact that, for a chronic disease such as CD, cumulative incidence over a long period provides a pragmatic registry-based spatial reference against which the PMSI-derived hospital-based signal can be compared. To place the two sources under a common standardisation framework, expected case counts were derived from the EPIMAD reference structure through indirect age- and sex-standardisation and used as a common offset for both datasets. This choice ensures that estimated discrepancies are not driven by source-specific definitions of expected counts.

We first describe the smoothed spatial patterns obtained separately for each source using source-specific Bayesian Poisson disease-mapping models with BYM2 spatial effects. We then compare the two databases jointly using the three proposed models: the random error model (REM), the structured error model (SEM), and the shared component model (SCM). In this setting, joint modelling is used to distinguish a possible global shift between sources from localised spatial discrepancies. Based on the simulation study, local disagreement was assessed primarily using the robustly centred exceedance probability (RCEP), with a tolerance band corresponding to a 10\% deviation on the relative-risk scale ($\varepsilon_{\mathrm{uni}}=\log(1.10)$), and declaring local disagreement when $\mathrm{RCEP}>0.9$.

\begin{figure}[htbp]
\centering
\includegraphics[width=\textwidth]{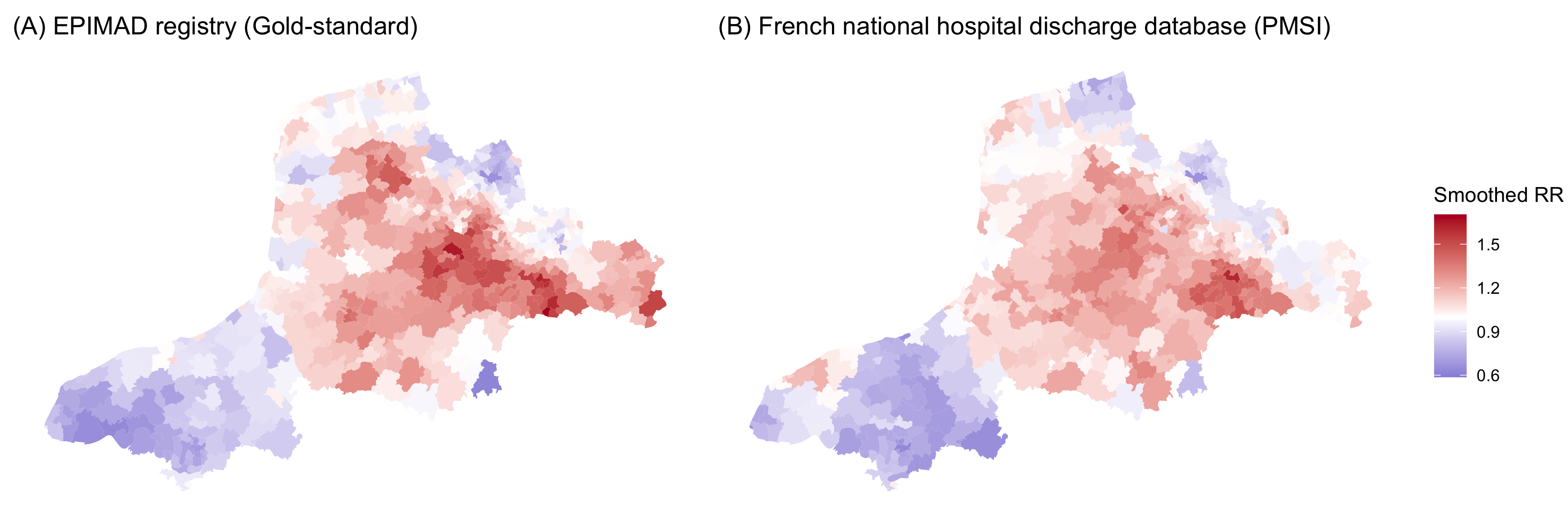}
\caption{Smoothed area-level relative risks for Crohn's disease obtained separately from the EPIMAD registry (A) and the French national hospital discharge database (PMSI; B) using source-specific Poisson models with BYM2 spatial effects. In the EPIMAD source, the mapped signal is based on cumulative incidence over 1988--2014, whereas in PMSI it is based on hospital-derived counts over 2007--2014. A common colour scale was used to facilitate direct visual comparison between the two sources.}
\label{fig:cd_maps_epimad_pmsi}
\end{figure}

Figure~\ref{fig:cd_maps_epimad_pmsi} displays the smoothed area-level relative risks obtained from these source-specific BYM2 models for EPIMAD and PMSI. The two sources exhibit broadly similar large-scale spatial gradients, with comparable areas of relatively high and low CD occurrence. Visual inspection nevertheless suggests a slightly lower overall intensity in PMSI than in EPIMAD, rather than a markedly different geographical organisation.

Joint modelling revealed a consistent global discrepancy between the two sources (REM: $RR_{\mathrm{global}}=0.932$ (95\% CrI: $[0.907;\,0.957]$); SEM:, $RR_{\mathrm{global}}=0.933$ (95\% CrI: $[0.907;\,0.959]$), and SCM $RR_{\mathrm{global}}=0.933$ (95\% CrI: $[0.906;\,0.960]$). Under the common EPIMAD-based expected-count structure, these results indicate that the PMSI-derived signal was, on average, about 7\% lower than the EPIMAD-based signal. The near-identical estimates obtained under the three model specifications further suggest that this global discrepancy is robust to assumptions regarding the structure of source-specific spatial variation.

Local discrepancy assessment based on RCEP did not identify any spatial unit exceeding the predefined detection threshold of 0.9, regardless of the joint model considered (REM with RCEP at 0.9: range from $1.59\times 10^{-6}$ to $3.67\times 10^{-6}$; median: $2.72\times 10^{-6}$; SCM with RCEP at 0.9: range from 0.101 to 0.638 (median: 0.243; interquartile range: 0.213--0.289); SEM with RCEP at 0.9: range from 0.094 to 0.849 (median: 0.296; interquartile range: 0.257--0.356). Although the spatial models produced broader posterior variability than REM, all local RCEP values remained below the detection threshold, and no local disagreement was identified. The corresponding area-level RCEP surfaces for the three fitted models are shown in Supplementary Figure~\ref{fig:sup_epimad_rcep}. No spatial unit exceeded the prespecified detection threshold.

Discrepancies between PMSI and EPIMAD are primarily global rather than spatially structured. Moreover, the PMSI-derived hospital-based signal appears to reproduce the broad geographical organisation of CD observed in the registry area, while underestimating its overall level. PMSI appears suitable for recovering the large-scale spatial geography of CD within the EPIMAD area, but less reliable for reproducing the absolute level of the registry-based reference signal without recalibration.

%

\section{Discussion}
\label{sec:discussion}

In this study, we proposed a Bayesian framework for comparing two spatial health-event databases when one source is treated as a reference and the other as a candidate reusable source. The main contribution of the work is to show that map-to-map validation should be understood through two complementary dimensions of disagreement in an paired and asymmetric manner: a \emph{global} discrepancy, summarised by the intercept contrast between sources, and a \emph{local} discrepancy, assessed through posterior decision rules applied to source-specific latent components.

A first key result is that the global intercept contrast behaved exactly according to its intended interpretation. Under the uniform perturbation scenario, $RR_{\mathrm{global}}$ accurately recovered the multiplicative shift across all three models. By contrast, under clustered or sparse perturbations, it remained much closer to 1 even when local differences were substantial. The intercept should only be interpreted as a summary of average map-wide displacement between the two databases. It evaluates the global calibration of the candidate source, but not whether it reproduces the same fine-scale spatial organisation.

Second, differences between models were generally small. REM and SEM behaved very similarly throughout the study, both for global and local summaries, suggesting that the main conclusions are not driven by whether the source-specific discrepancy field is represented as unstructured or spatially structured residual variation. SCM showed a more distinctive profile and was often more conservative in weak one-sided alternatives, particularly with RCEP. This is coherent with its formulation: unlike REM and SEM, the source-specific component in SCM is not a direct residual contrast with the reference source, but rather a source-specific deviation conditional on a shared spatial field. SCM therefore remains useful when the objective is to partition common and source-specific spatial structure in a more symmetric way \citep{knorr-held_shared_2001,retegui_estimating_2021,etxeberria_using_2023}. However, for the specific purpose of validating a candidate database against a reference source, the error-model family appears conceptually more natural because the asymmetry between sources is made explicit in the model itself.

The third main result concerns local discrepancy detection. Across the simulation study, the RCEP generally provided the most stable and interpretable operating characteristics. Its advantage over the NREP was especially clear in clustered and sparse random settings, where it provided the best trade-off between sensitivity and false-positive control, and therefore the highest MCC values. This is consistent with the modelling rationale: because discrepancy fields are embedded in centred latent models, direct thresholding around 0 may partly reflect the compensatory structure induced by the latent field rather than genuinely unusual local departures. By re-centring inference around a robust estimate of the overall discrepancy level, RCEP is better able to identify areas that are atypical relative to the discrepancy surface as a whole. NREP was sometimes more sensitive in weak one-sided clustered settings, but this advantage generally came at the cost of lower specificity and higher false discovery rates. In real-data applications, however, the form of local discrepancy is unknown in advance, and one cannot assume that departures, if present, will correspond to the narrow settings in which NREP may gain sensitivity. From that perspective, RCEP appears to be the most appropriate default rule for local discrepancy detection, as it offers more reliable behaviour across the range of perturbation structures considered while better controlling spurious detections. In the settings explored here, threshold 0.9 provided the most convincing overall compromise.

The EPIMAD--PMSI application illustrates the distinction between global and local disagreement particularly clearly. All three joint models led to the same conclusion: the PMSI-derived signal was globally lower than the EPIMAD-based signal by about 7\%, while no spatial unit exceeded the local RCEP detection threshold. Under the common EPIMAD-based expected-count structure, this indicates that PMSI reproduces the broad geographical organisation of Crohn's disease observed in the registry area reasonably well, but underestimates its overall level. This is substantively important for data reuse in spatial epidemiology, because it suggests that medico-administrative data may preserve the large-scale spatial gradients relevant for environmental or territorial analyses, while still requiring recalibration if the objective is to recover the absolute level of the registry-based reference signal.

Several limitations should be acknowledged. First, the simulation study was conducted on a single areal configuration and within a controlled simulation framework, which may limit the generalisability of the conclusions. Supplementary sensitivity analyses varying baseline incidence and spatial correlation confirmed the relative robustness of the global discrepancy summary and the greater fragility of local detection in rare-event settings. Additional analyses on the RCEP tolerance parameter also showed that the main local conclusions remained stable under reasonable choices of equivalence-band width. Further work would nevertheless be useful to explore more extensively the roles of spatial variance, neighbourhood structure, and prior specification. Second, only one family of prior specifications was explored in the main analyses. While the regularising PC priors adopted here are well motivated and consistent with current BYM2 practice \citep{riebler_pcprior_2016}, some aspects of performance, especially for SCM, may depend on prior choice and identifiability constraints. Third, RCEP is a plug-in posterior functional, since the robust centre is estimated and then treated as fixed; this is operationally convenient, but it does not propagate the full uncertainty attached to that centring step.

A further limitation concerns the nature of the quantities compared in the application. EPIMAD provides registry-based incident cases, whereas PMSI-derived counts are based on hospital stays and more plausibly reflect hospital-based disease burden among patients treated for Crohn's disease than incidence itself. The application should therefore be interpreted primarily as a validation of spatial signal rather than as a strict comparison of identical epidemiological estimands. More broadly, this example illustrates a common challenge in health-data reuse: external databases may reproduce relevant spatial structure even when they do not measure exactly the same underlying quantity as the reference source.

The proposed framework should not be viewed as intrinsically restricted to Poisson disease-mapping models. Its underlying logic, namely the joint comparison of a reference source and a candidate reusable source through separate global and local discrepancy summaries, could be extended to other outcome distributions and likelihood structures depending on the nature of the data, including binomial, negative binomial, Gaussian, or survival-type models. Likewise, extending the framework to the spatio-temporal setting is a natural next step, as it would make it possible to distinguish overall temporal shifts, persistent local discrepancies, and transient area-specific divergences between sources. As the French national cancer registry is being established, calibrating a model in areas with cancer registries could be seen as a promising method to determine the risk of cancer in uncovered areas.

Overall, the present results support a simple practical recommendation for map-to-map spatial validation. The global intercept contrast should be used systematically to assess whether the candidate source is globally shifted relative to the reference database. For local discrepancy detection, RCEP appears to be the most appropriate default decision rule, while NREP may still be useful when the inferential target is strict departure from the null reference value. More fundamentally, the proposed framework helps avoid a common interpretative pitfall: concluding that two maps ``agree'' merely because no local discrepancy is detected, while overlooking a substantial territory-wide shift. In that sense, the method provides not only a modelling strategy, but also a clearer conceptual basis for the validation and reuse of spatial health databases.

\bibliographystyle{apalike}
\bibliography{biblio_harmonized_publication_ready}


\clearpage
\appendix
\setcounter{figure}{0}
\setcounter{table}{0}
\renewcommand{\thesection}{S\arabic{section}}
\renewcommand{\thetable}{S\arabic{table}}
\renewcommand{\thefigure}{S\arabic{figure}}

\section{Supplementary materials}
\label{supp_mat}

For each simulated dataset, local detection performance was evaluated by comparing the true discrepancy status of each area with the label predicted by the posterior decision rule. Let $Y_i$ denote the true status of area $i$, with $Y_i=1$ if area $i$ was truly perturbed and $Y_i=0$ otherwise. Let $\widehat{Y}_i$ denote the corresponding predicted status, with $\widehat{Y}_i=1$ if area $i$ was flagged as discrepant by the decision rule and $\widehat{Y}_i=0$ otherwise. The resulting classification table was defined as follows:
\[
\begin{array}{c|cc}
 & Y_i=1 & Y_i=0 \\
\midrule
\widehat{Y}_i=1 & \mathrm{TP} & \mathrm{FP} \\
\widehat{Y}_i=0 & \mathrm{FN} & \mathrm{TN}
\end{array}
\]
where TP, TN, FP, and FN denote the numbers of true positives, true negatives, false positives, and false negatives, respectively. Sensitivity, specificity, and false discovery rate were defined as
\[
\mathrm{Sensitivity}
=
\frac{\mathrm{TP}}{\mathrm{TP}+\mathrm{FN}},
\qquad
\mathrm{Specificity}
=
\frac{\mathrm{TN}}{\mathrm{TN}+\mathrm{FP}},
\qquad
\mathrm{FDR}
=
\frac{\mathrm{FP}}{\mathrm{TP}+\mathrm{FP}}.
\]

Overall classification performance was summarised using the Matthews correlation coefficient (MCC), defined as
\[
\mathrm{MCC}
=
\frac{
\mathrm{TP}\times\mathrm{TN}
-
\mathrm{FP}\times\mathrm{FN}
}{
\sqrt{
(\mathrm{TP}+\mathrm{FP})
(\mathrm{TP}+\mathrm{FN})
(\mathrm{TN}+\mathrm{FP})
(\mathrm{TN}+\mathrm{FN})
}
}.
\]

Metrics with a zero denominator, for example sensitivity when no truly discrepant areas were present, specificity when no truly non-discrepant areas were present, or FDR when no area was flagged, were treated as not applicable rather than set to zero.

\newpage

\begin{table}[ht]
\centering
\begin{tabular}{cp{3cm}p{12cm}}
\toprule
\textbf{Scenario} & \textbf{Name} & \textbf{Description} \\
\midrule
S1 & Null (no perturbation) & No area is modified. All perturbation factors are equal to 1: $r_i = 1$ for all $i = 1,\dots,I$. \\
\midrule
S2 & Uniform disturbance & All areas are modified by the same multiplicative factor: $r_i = r$ for all $i = 1,\dots,I$, with $r \in \{0.25, 0.50, 0.75, 1.25, 1.5, 1.75, 2.0, 2.25, 2.5\}$. \\
\midrule
S3 & Clustered disturbance & Perturbations are restricted to a fixed connected spatial cluster. In one-sided settings, all areas in the cluster receive the same factor $r$, while others remain unchanged. In mixed settings, two disjoint clusters are perturbed in opposite directions (one upward, one downward). \\
\midrule
S4 & Random disturbance & A proportion $\pi \in \{0.25, 0.50\}$ of areas is randomly selected without replacement. In one-sided settings, selected areas receive the same factor $r$. In mixed settings, each selected area is independently assigned an upward or downward perturbation with probability 0.5. Mixed perturbations were also considered using pairs $(1.5, 0.75)$ and $(1.75, 0.50)$. \\
\bottomrule
\end{tabular}
\captionsetup{justification=raggedright, singlelinecheck=false}
\caption{Definition of simulation scenarios based on spatial perturbation structures. 
$r_i$ denotes the multiplicative perturbation factor applied to area $i$ ($i=1,\dots,I$), where $I$ is the total number of spatial units. 
The indicator $Y_i$ equals 1 if area $i$ is perturbed and 0 otherwise. In mixed settings, both upward and downward perturbations are considered as modified areas ($Y_i=1$).}
\label{tab:simulation_scenarios}
\end{table}

\newpage

\begin{figure}[htbp]
\begin{center}
\includegraphics[width=0.7\textwidth]{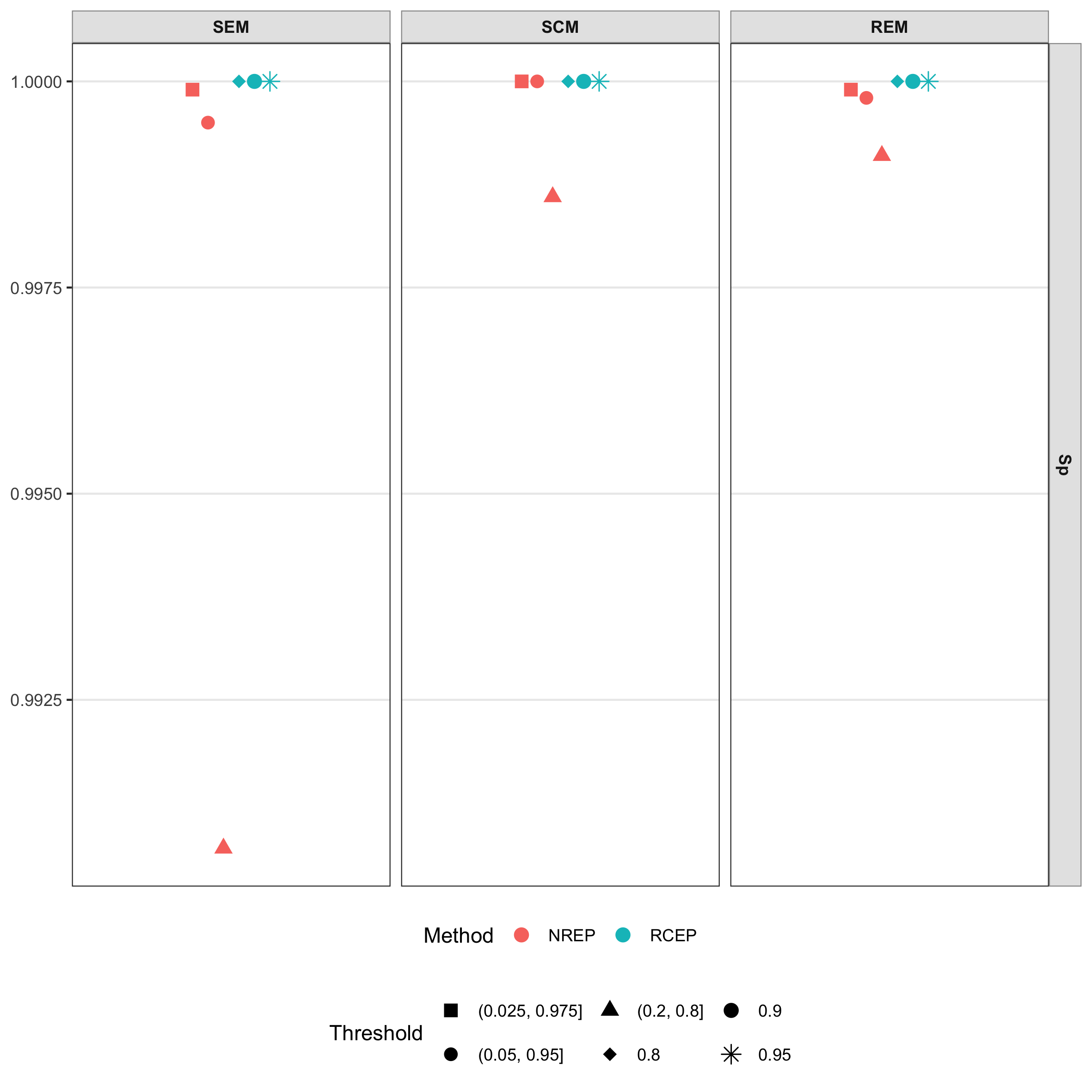}
\caption{Classification performance in Scenario~S1 (null scenario), where no local difference is present between the two maps. Only specificity (Sp) is displayed, since sensitivity is not defined in the absence of truly perturbed areas. Results are shown for the three joint models (SEM, SCM, and REM), comparing the null-reference exceedance probability approach (NREP) and the robustly centred exceedance probability approach (RCEP) across decision thresholds. Values closer to 1 indicate better control of false positive detections.}
\label{fig:sup_Sc1}
\end{center}
\end{figure}

\begin{figure}[htbp]
\begin{center}
\includegraphics[width=\textwidth]{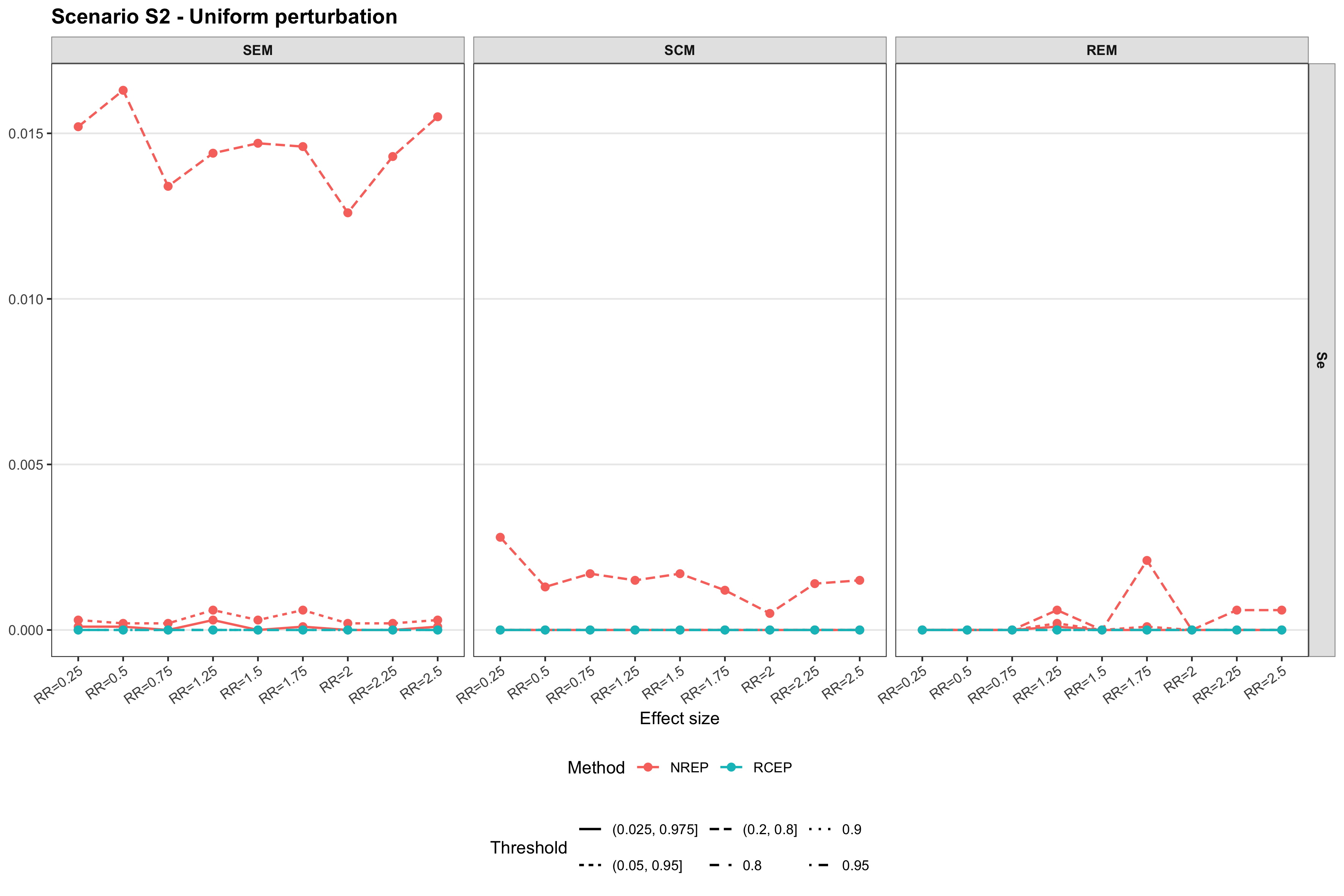}
\caption{Classification performance in Scenario~S2 (uniform perturbation), where the two maps differ only through a global shift affecting all areas equally. Only sensitivity (Se) is displayed, because all areas are perturbed and specificity is therefore not defined. Results are shown for the three joint models (SEM, SCM, and REM), comparing NREP and RCEP across decision thresholds and perturbation magnitudes. This figure highlights the ability, or inability, of local detection rules to recover a purely global signal.}
\label{fig:sup_Sc2}
\end{center}
\end{figure}

\begin{landscape}
\begin{figure}[htbp]
\begin{center}
\includegraphics[width=\textwidth]{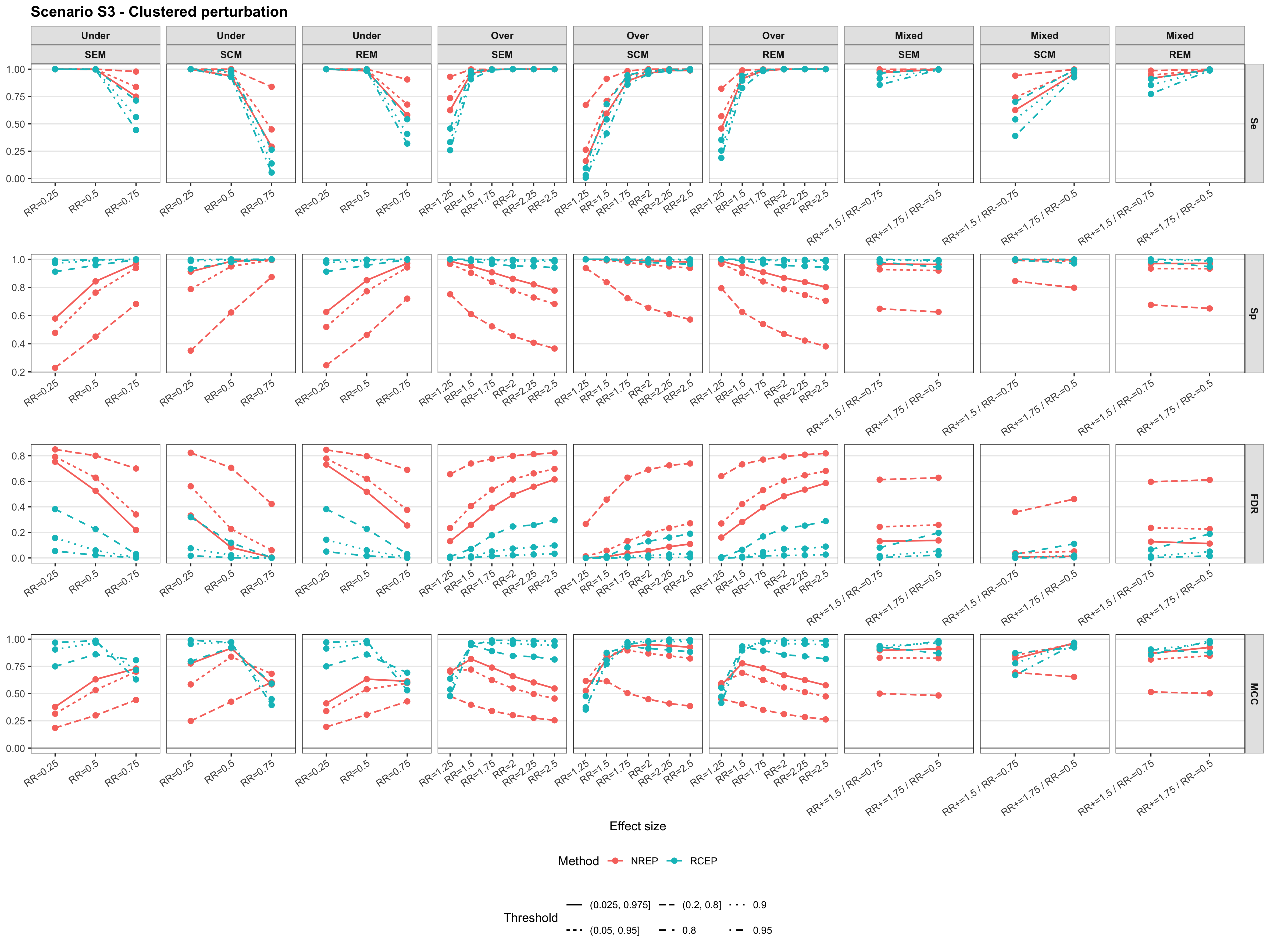}
\caption{Classification performance in Scenario~S3 (clustered perturbation), where differences between the two maps are spatially structured and localized within a cluster of areas. Sensitivity (Se), specificity (Sp), false discovery rate (FDR), and Matthews correlation coefficient (MCC) are displayed for the three joint models (SEM, SCM, and REM), separately for under-incidence, over-incidence, and mixed perturbation settings. Results compare the null-reference exceedance probability approach (NREP) and the robustly centred exceedance probability approach (RCEP) across decision thresholds. Higher Se, Sp, and MCC, together with lower FDR, indicate better classification performance.}
\label{fig:sup_Sc3}
\end{center}
\end{figure}

\end{landscape}

\begin{landscape}
\begin{figure}[htbp]
\begin{center}
\includegraphics[width=\textwidth]{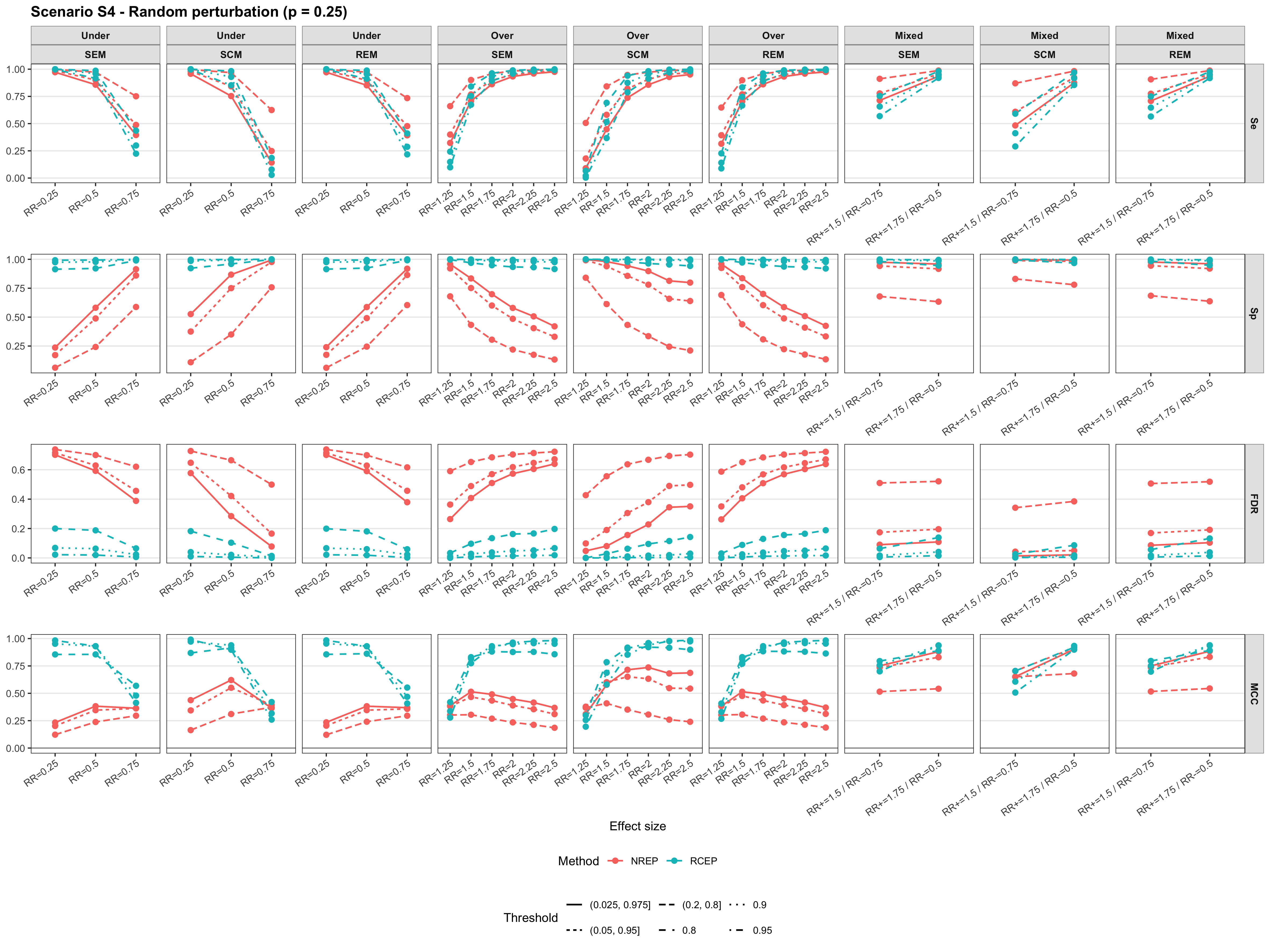}
\caption{Classification performance in Scenario~S4 (random perturbation) with $p=0.25$, meaning that 25\% of areas are perturbed at random. Sensitivity (Se), specificity (Sp), false discovery rate (FDR), and Matthews correlation coefficient (MCC) are displayed for the three joint models (SEM, SCM, and REM), separately for under-incidence, over-incidence, and mixed perturbation settings. Results compare NREP and RCEP across decision thresholds and perturbation magnitudes. This figure summarizes performance when local differences are sparse and spatially unstructured.}
\label{fig:sup_Sc4_025}
\end{center}
\end{figure}
\end{landscape}

\begin{landscape}
\begin{figure}[htbp]
\begin{center}
\includegraphics[width=\textwidth]{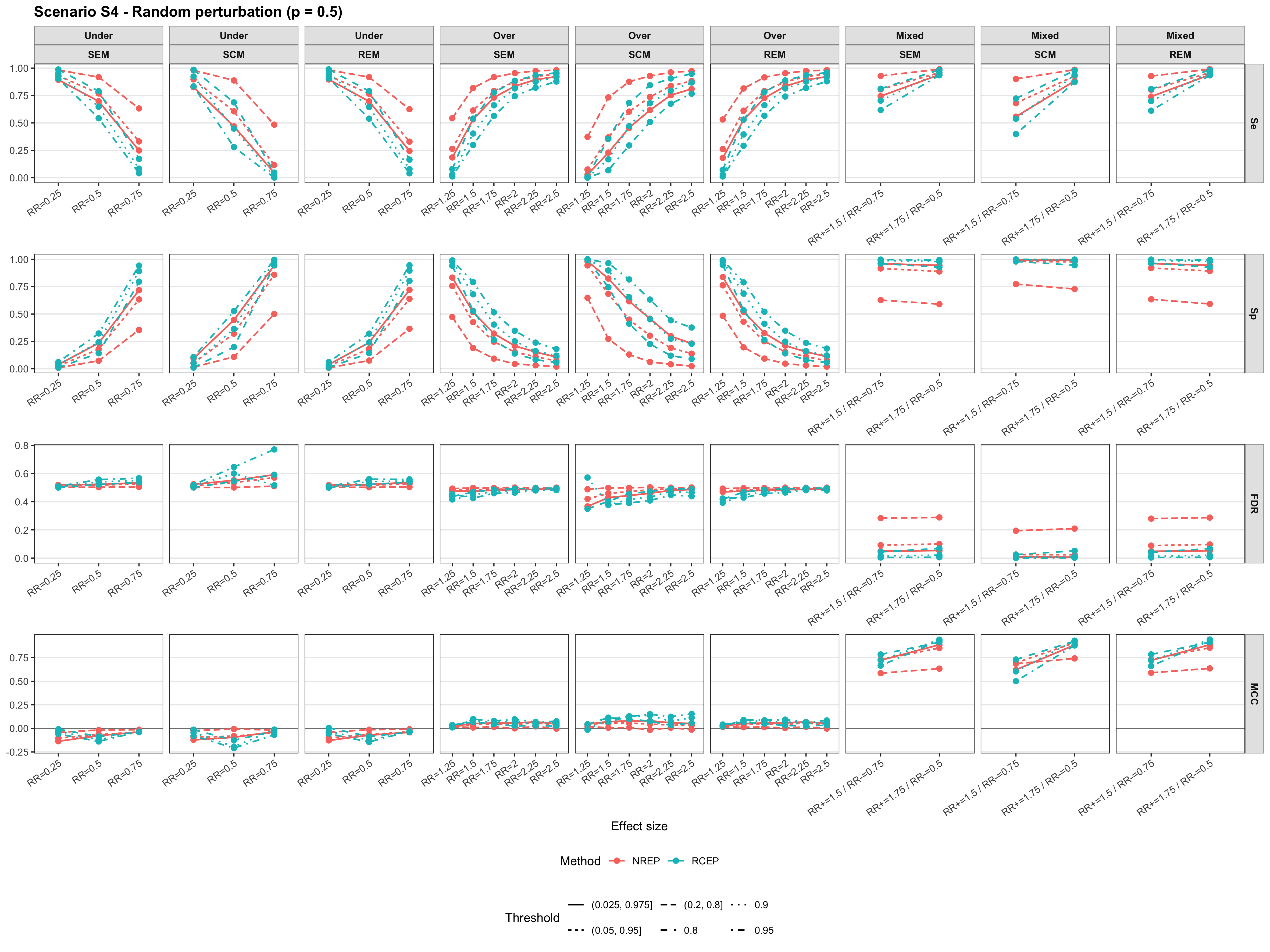}
\caption{Classification performance in Scenario~S4 (random perturbation) with $p=0.50$, meaning that 50\% of areas are perturbed at random. Sensitivity (Se), specificity (Sp), false discovery rate (FDR), and Matthews correlation coefficient (MCC) are displayed for the three joint models (SEM, SCM, and REM), separately for under-incidence, over-incidence, and mixed perturbation settings. Results compare NREP and RCEP across decision thresholds and perturbation magnitudes. Relative to the $p=0.25$ setting, this figure illustrates performance when perturbations affect a larger proportion of the study area.}
\label{fig:sup_Sc4_05}
\end{center}
\end{figure}
\end{landscape}

\begin{figure}[htbp]
\begin{center}
\includegraphics[width=0.8\textwidth]{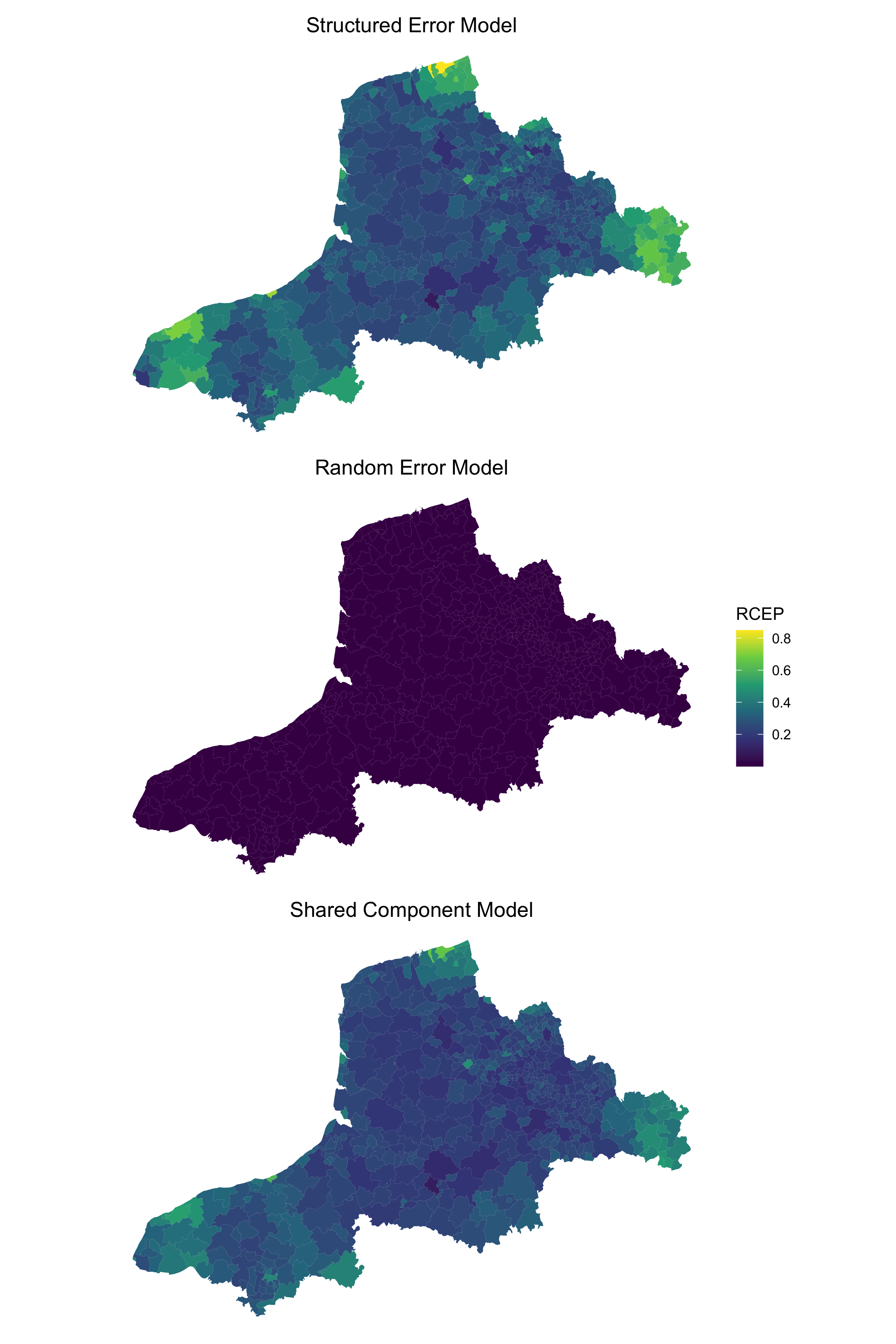}
\caption{Area-level robustly centred exceedance probabilities (RCEP) for the EPIMAD-PMSI comparison under the structured error model (SEM), random error model (REM), and shared component model (SCM). These maps show the spatial distribution of posterior RCEP values for each fitted model. Although SEM and SCM display broader variation than REM, all RCEP values remain below the prespecified detection threshold of 0.9, indicating that no local disagreement was identified.}
\label{fig:sup_epimad_rcep}
\end{center}
\end{figure}

\begin{figure}[htbp]
\begin{center}
\includegraphics[width=0.85\textwidth]{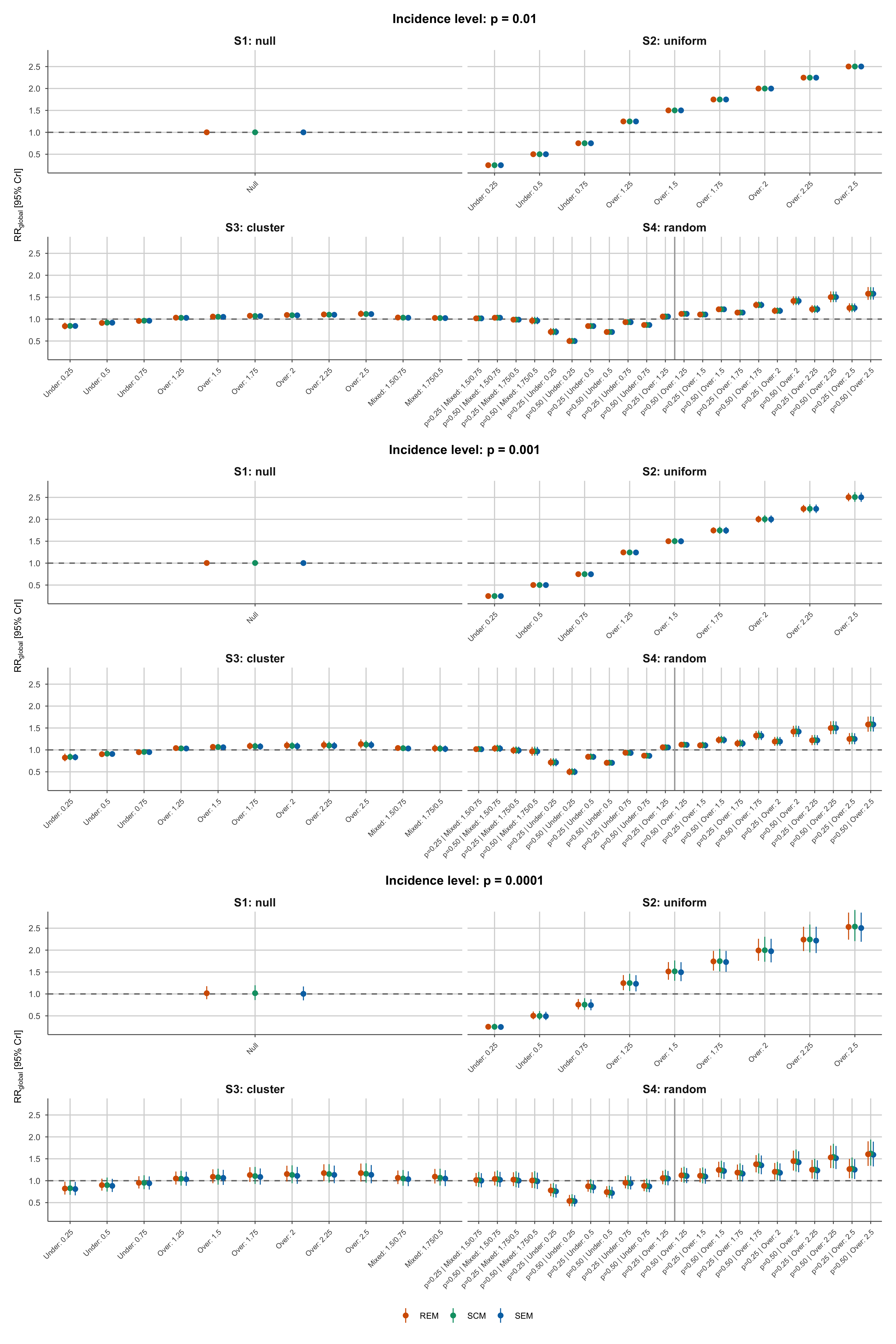}
\caption{Posterior median of the global relative-risk contrast, $RR_{\mathrm{global}}=\exp(\Delta)$, across simulation scenarios and perturbation settings, shown separately for three baseline incidence levels ($p=0.01$, $0.001$, and $0.0001$). Results are displayed for the random error model (REM), shared component model (SCM), and structured error model (SEM), together with corresponding 95\% credible intervals. As incidence decreases, the global discrepancy summary remains broadly well calibrated and continues to recover map-wide shifts, although uncertainty increases at the rarest incidence level.}
\label{fig:sup_incidence_global}
\end{center}
\end{figure}

%

\begin{figure}[htbp]
\begin{center}
\includegraphics[width=\textwidth]{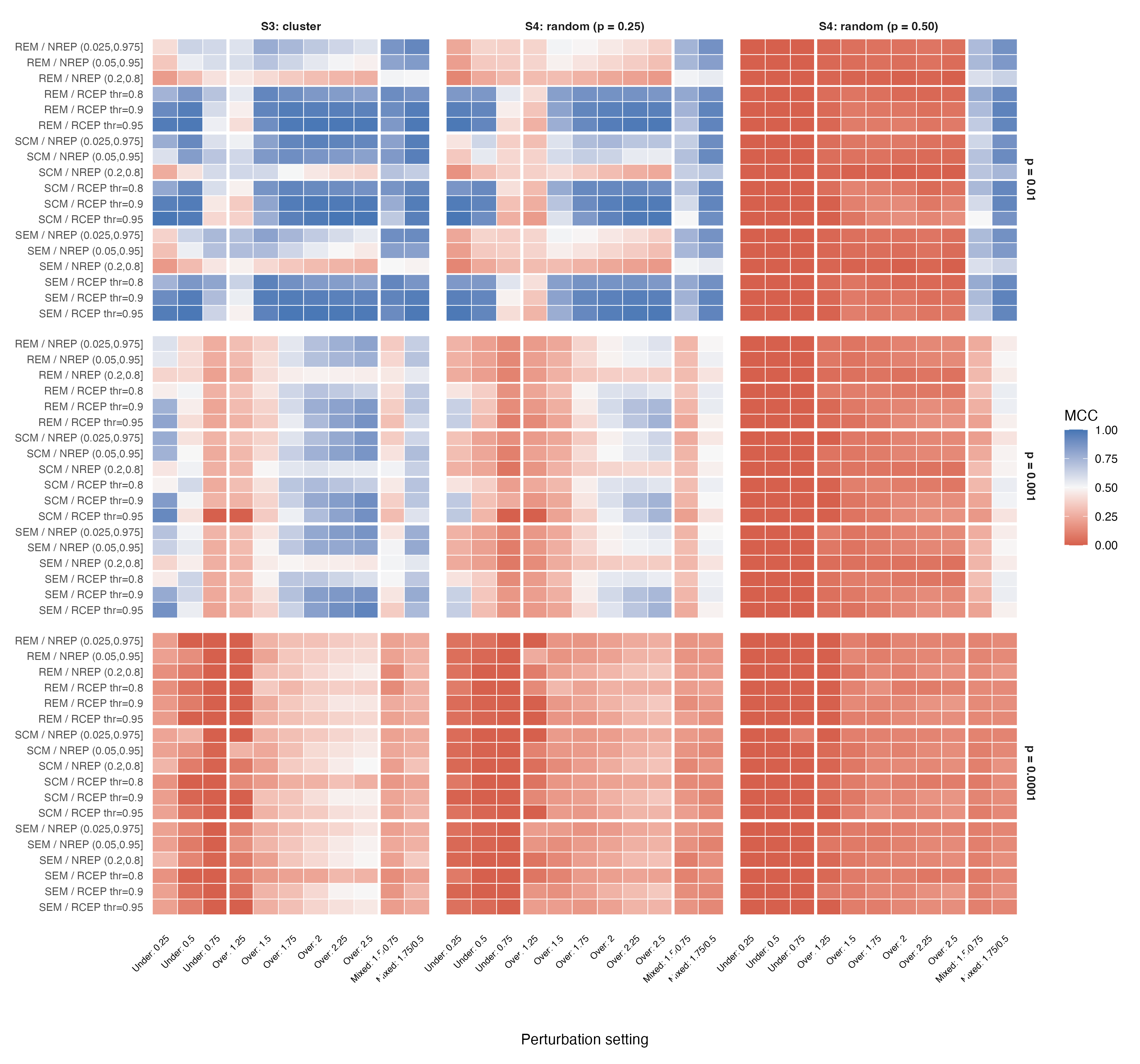}
\caption{Mean Matthews correlation coefficient (MCC) across simulation settings for clustered perturbations (S3) and random perturbations (S4), shown separately for three baseline incidence levels ($p=0.01$, $0.001$, and $0.0001$). Results are displayed for the REM, SCM, and SEM under the null-referenced exceedance probability (NREP) and the robustly centred exceedance probability (RCEP). Higher values indicate better overall local classification performance. This figure highlights the marked deterioration of local discrepancy detection as incidence decreases, especially in weak or spatially heterogeneous perturbation settings.}
\label{fig:sup_incidence_mcc}
\end{center}
\end{figure}

%

\begin{figure}[htbp]
\begin{center}
\includegraphics[width=0.85\textwidth]{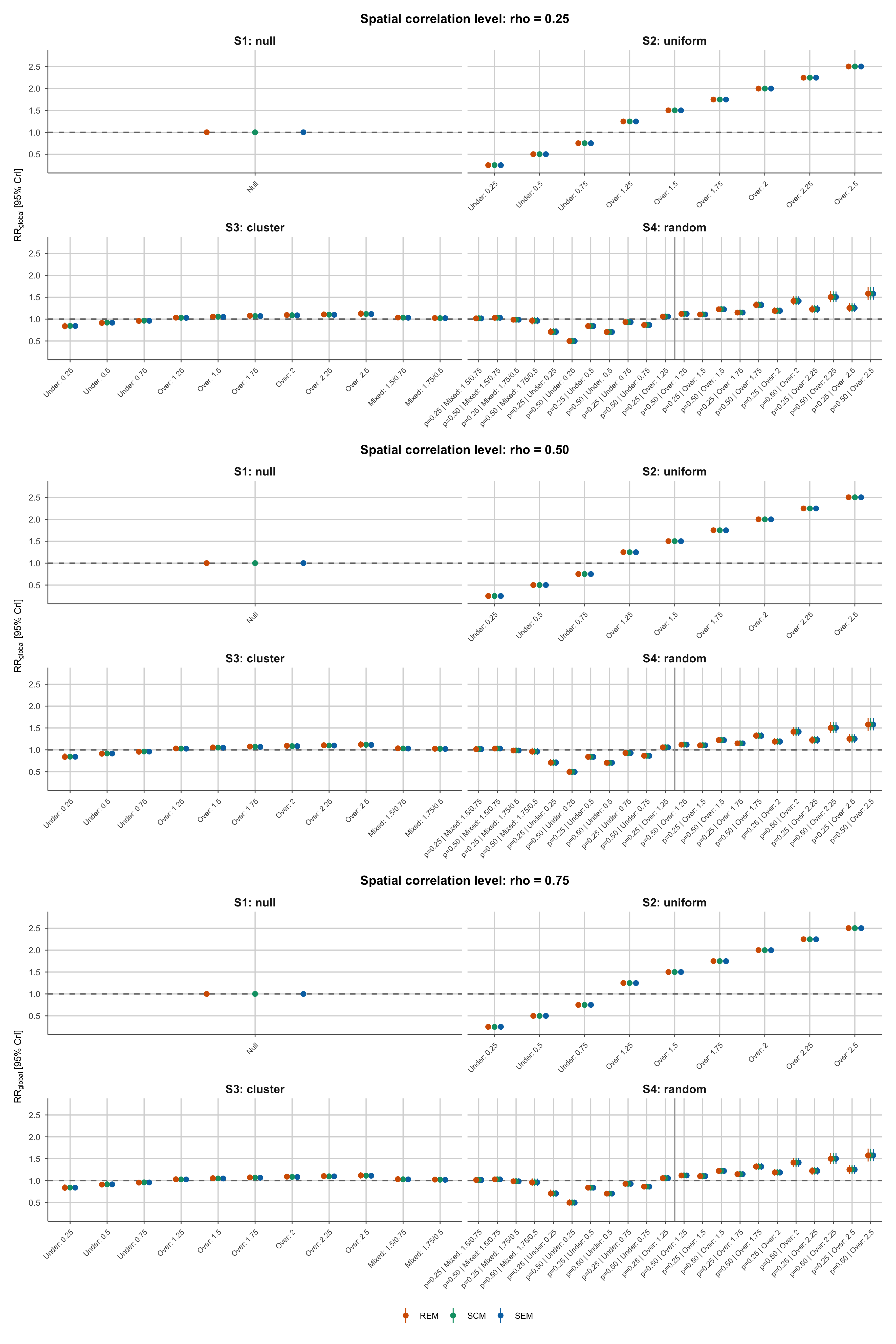}
\caption{Posterior median of the global relative-risk contrast, $RR_{\mathrm{global}}=\exp(\Delta)$, across simulation scenarios and perturbation settings, shown separately for three spatial correlation levels of the latent Leroux field ($\rho=0.25$, $0.50$, and $0.75$). Results are displayed for the REM, SCM, and SEM, together with corresponding 95\% credible intervals. The global discrepancy summary is essentially unchanged across correlation levels, indicating strong robustness of the intercept contrast to the strength of latent spatial dependence.}
\label{fig:sup_rho_global}
\end{center}
\end{figure}

%

\begin{figure}[htbp]
\begin{center}
\includegraphics[width=\textwidth]{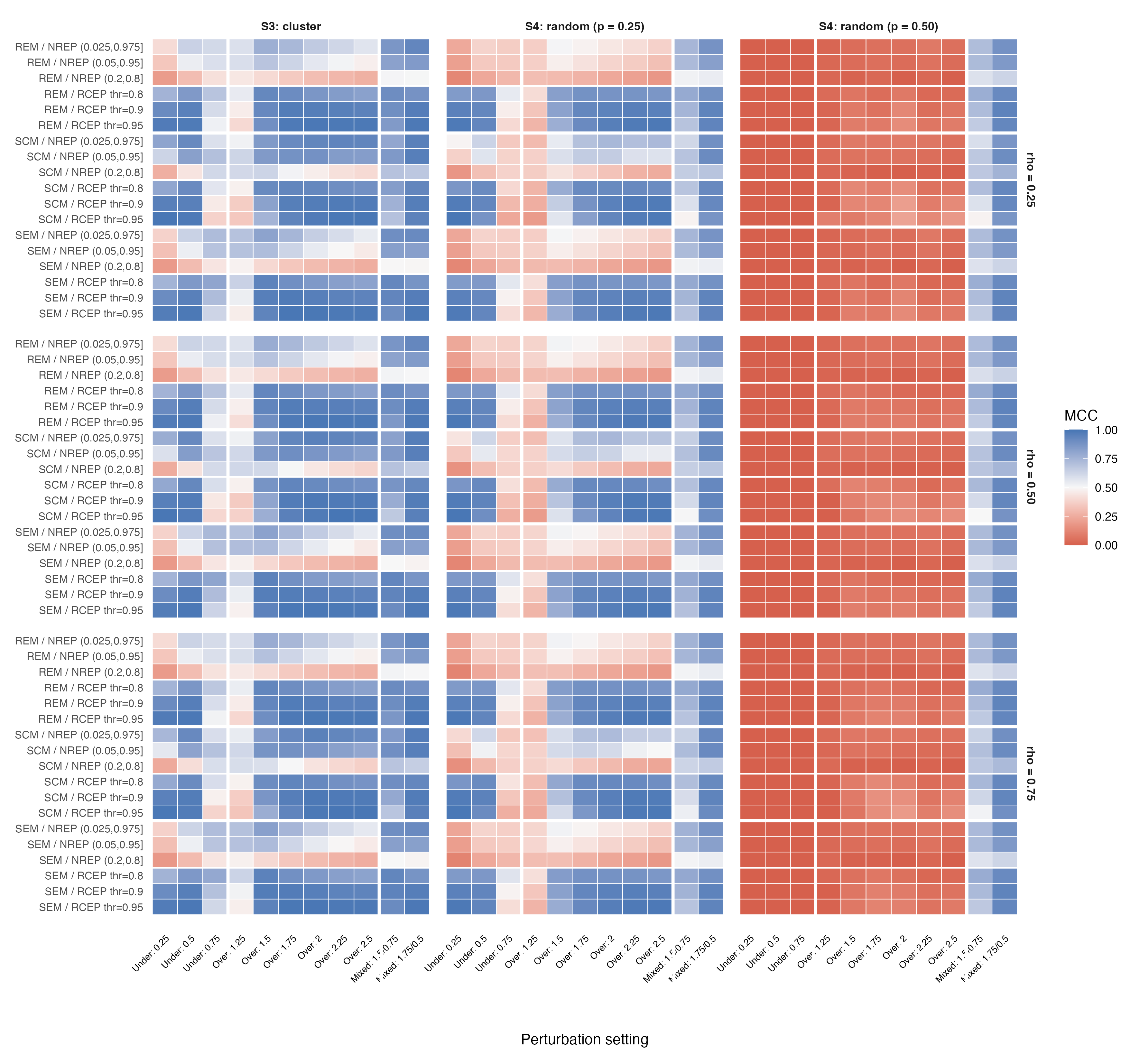}
\caption{Mean Matthews correlation coefficient (MCC) across simulation settings for clustered perturbations (S3) and random perturbations (S4), shown separately for three spatial correlation levels of the latent Leroux field ($\rho=0.25$, $0.50$, and $0.75$). Results are displayed for the REM, SCM, and SEM under the null-referenced exceedance probability (NREP) and the robustly centred exceedance probability (RCEP). Higher values indicate better overall local classification performance. This figure shows that varying the spatial correlation strength has only modest effects on local detection performance and does not alter the overall ranking of models or posterior decision rules.}
\label{fig:sup_rho_mcc}
\end{center}
\end{figure}

\begin{figure}[htbp]
\begin{center}
\includegraphics[width=\textwidth]{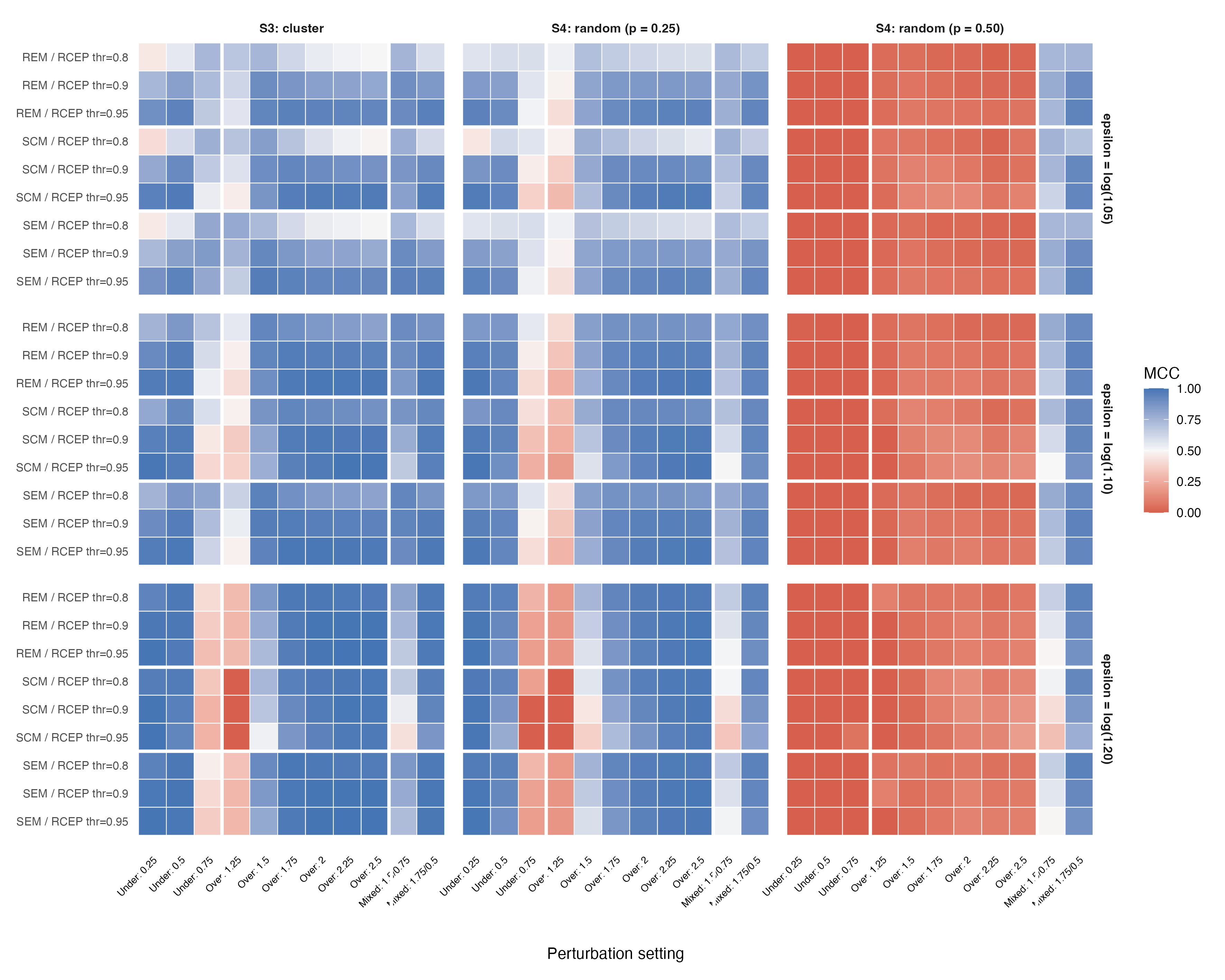}
\caption{Mean Matthews correlation coefficient (MCC) across simulation settings for clustered perturbations (S3) and random perturbations (S4), shown separately for three equivalence-band widths used in the robustly centred exceedance probability ($\epsilon=\log(1.05)$, $\log(1.10)$, and $\log(1.20)$). Results are displayed for the REM, SCM, and SEM under RCEP only, with thresholds 0.8, 0.9, and 0.95. Higher values indicate better overall local classification performance. This figure illustrates the expected trade-off induced by $\epsilon$: narrower equivalence bands yield more liberal detection, whereas wider bands are more conservative, with $\epsilon=\log(1.10)$ providing the most balanced overall behaviour.}
\label{fig:sup_eps_mcc}
\end{center}
\end{figure}


\end{document}